\documentclass[aps,prx,floatfix, linenumbers,twocolumn,superscriptaddress,tightenlines,nobibnotes,amsmath,amssymb]{revtex4}

\usepackage{graphicx}
\usepackage{amsmath}
\usepackage{amssymb}
\usepackage{mathrsfs}
\usepackage{dsfont}
\usepackage{epstopdf}
\usepackage{color}
\usepackage{natbib}

\newcommand{\bra}[1] {\langle #1|}
\newcommand{\ket}[1] {|#1 \rangle}
\newcommand{\braket}[2] {\langle #1|#2 \rangle}

\newcommand*{\rom}[1]{\expandafter\@slowromancap\romannumeral #1@}

\begin{document}

\title{Flopping-mode electric dipole spin resonance in phosphorus donor qubits in silicon}
\author{F. N. Krauth}
\affiliation{Centre of Excellence for Quantum Computation and Communication Technology, School of Physics, University of New South Wales, Sydney, New South Wales 2052, Australia}
\affiliation{Silicon Quantum Computing Pty Ltd., Level 2, Newton Building, UNSW Sydney, Kensington, NSW 2052, Australia}
\author{S. K. Gorman}
\affiliation{Centre of Excellence for Quantum Computation and Communication Technology, School of Physics, University of New South Wales, Sydney, New South Wales 2052, Australia}
\affiliation{Silicon Quantum Computing Pty Ltd., Level 2, Newton Building, UNSW Sydney, Kensington, NSW 2052, Australia}
\author{Y. He}
\affiliation{Centre of Excellence for Quantum Computation and Communication Technology, School of Physics, University of New South Wales, Sydney, New South Wales 2052, Australia}
\affiliation{Silicon Quantum Computing Pty Ltd., Level 2, Newton Building, UNSW Sydney, Kensington, NSW 2052, Australia}
\author{M. T. Jones}
\affiliation{Centre of Excellence for Quantum Computation and Communication Technology, School of Physics, University of New South Wales, Sydney, New South Wales 2052, Australia}
\affiliation{Silicon Quantum Computing Pty Ltd., Level 2, Newton Building, UNSW Sydney, Kensington, NSW 2052, Australia}
\author{P. Macha}
\affiliation{Centre of Excellence for Quantum Computation and Communication Technology, School of Physics, University of New South Wales, Sydney, New South Wales 2052, Australia}
\affiliation{Silicon Quantum Computing Pty Ltd., Level 2, Newton Building, UNSW Sydney, Kensington, NSW 2052, Australia}
\author{S. Kocsis}
\affiliation{Silicon Quantum Computing Pty Ltd., Level 2, Newton Building, UNSW Sydney, Kensington, NSW 2052, Australia}
\affiliation{School of Physics, University of New South Wales, Sydney, New South Wales 2052, Australia}
\author{C. Chua}
\affiliation{Silicon Quantum Computing Pty Ltd., Level 2, Newton Building, UNSW Sydney, Kensington, NSW 2052, Australia}
\affiliation{School of Physics, University of New South Wales, Sydney, New South Wales 2052, Australia}
\author{B. Voisin}
\affiliation{Silicon Quantum Computing Pty Ltd., Level 2, Newton Building, UNSW Sydney, Kensington, NSW 2052, Australia}
\affiliation{School of Physics, University of New South Wales, Sydney, New South Wales 2052, Australia}
\author{S. Rogge}
\affiliation{Centre of Excellence for Quantum Computation and Communication Technology, School of Physics, University of New South Wales, Sydney, New South Wales 2052, Australia}
\author{R. Rahman}
\affiliation{Silicon Quantum Computing Pty Ltd., Level 2, Newton Building, UNSW Sydney, Kensington, NSW 2052, Australia}
\affiliation{School of Physics, University of New South Wales, Sydney, New South Wales 2052, Australia}
\author{Y. Chung}
\affiliation{Centre of Excellence for Quantum Computation and Communication Technology, School of Physics, University of New South Wales, Sydney, New South Wales 2052, Australia}
\affiliation{Silicon Quantum Computing Pty Ltd., Level 2, Newton Building, UNSW Sydney, Kensington, NSW 2052, Australia}
\author{M. Y. Simmons}
\affiliation{Centre of Excellence for Quantum Computation and Communication Technology, School of Physics, University of New South Wales, Sydney, New South Wales 2052, Australia}
\affiliation{Silicon Quantum Computing Pty Ltd., Level 2, Newton Building, UNSW Sydney, Kensington, NSW 2052, Australia}

\date{\today}

\begin{abstract}
Single spin qubits based on phosphorus donors in silicon are a promising candidate for a large-scale quantum computer. Despite long coherence times, achieving uniform magnetic control remains a hurdle for scale-up due to challenges in high-frequency magnetic field control at the nanometre-scale. Here, we present a proposal for a flopping-mode electric dipole spin resonance qubit based on the combined electron and nuclear spin states of a double phosphorus donor quantum dot. The key advantage of utilising a donor-based system is that we can engineer the number of donor nuclei in each quantum dot. By creating multi-donor dots with antiparallel nuclear spin states and multi-electron occupation we can minimise the longitudinal magnetic field gradient, known to couple charge noise into the device and dephase the qubit. We describe the operation of the qubit and show that by minimising the hyperfine interaction of the nuclear spins we can achieve $\pi/2-X$ gate error rates of $\sim10^{-4}$ using realistic noise models. We highlight that the low charge noise environment in these all-epitaxial phosphorus-doped silicon qubits will facilitate the realisation of strong coupling of the qubit to superconducting microwave cavities allowing for long-distance two-qubit operations.
\end{abstract}

\maketitle


Electron spin resonance (ESR) using high-frequency magnetic fields allows for high-fidelity single-qubit ($F > 99$ \%) gates in donor-based silicon qubits~\cite{Muhonen2014}. The technical complexity of generating local oscillating magnetic fields at nanometre length scales in semiconductor qubits however, remains a challenge for the future scalability of magnetic control~\cite{vahapoglu2021}. The tight-packing in exchange-based spin qubits, in which donors are only a few tens of nanometres apart, creates a challenge in minimising crosstalk between them~\cite{Buch2013}. As a result, there has been a growing interest in electric dipole spin resonance (EDSR) to electrically control qubits with local electric fields and coupling the qubits via their charge dipole moment. Electric dipole spin resonance is achieved by coupling the spin of an electron to its charge degree-of-freedom allowing the spin state to be controlled by moving the electron using electric fields~\cite{PhysRevB.78.195302}. This spin-charge coupling can be created by a number of different physical mechanisms such as the use of large spin-orbit coupling materials~\cite{PhysRevB.67.115324, Nowack1430, NadjPerge2010}, magnetic field gradients from micromagnets~\cite{Pioro2007, PhysRevLett.96.047202, Yoneda2018, PhysRevResearch.2.012006}, and the hyperfine interaction between the electron and surrounding nuclear spins~\cite{PhysRevLett.99.246601, PhysRevLett.110.107601, Tosi2017}.

Depending on the nature of the physical mechanism that couples the spin and charge degree-of-freedom there are also various differences in the way EDSR can be used to drive qubit operations. The use of materials with intrinsic spin-orbit coupling such as III-V semiconductor materials~\cite{PhysRevB.67.115324, Nowack1430, NadjPerge2010, Petersson2012} allows for EDSR without the need of any additional control structures~\cite{Pioro2007}. For material systems with low intrinsic spin-orbit coupling such as electrons in silicon it is difficult to operate a qubit using EDSR without creating spin-orbit coupling using extrinsic mechanisms. To generate a synthetic spin-orbit coupling, micromagnets were therefore introduced to create a gradient magnetic field near the spin qubits~\cite{Pioro2007}. However, these micromagnets not only require further processing steps, complicating device architectures, but have also been shown to introduce additional charge noise~\cite{Kha2015}. When an electron is moved back-and-forth within the magnetic field gradient perpendicular to the static magnetic field, $B_0$, it experiences an effective oscillating magnetic field with a corresponding energy, $\Delta\Omega_{\perp}$ which can be used to drive spin rotations~\cite{Yoneda2018}. However, any stray magnetic field gradient parallel to $B_0$ with a corresponding energy, $\Delta\Omega_{\parallel}$ leads to charge noise induced dephasing. In this manuscript, we consider flopping-mode EDSR where a single electron is shuttled between two donor-based quantum dots~\cite{Hu2012,Beaudoin2016,PhysRevResearch.2.012006} rather than shaking an electron within a single quantum dot~\cite{Nowack1430}. The proposed qubit is shown to achieve long coherence times by reducing the longitudinal magnetic field gradient while maintaining a large $\sim 100$ MHz transverse magnetic field gradient. Additionally, these flopping-mode qubits can then be measured via dispersive charge readout~\cite{PhysRevResearch.2.012006} or by direct single-shot spin readout~\cite{Elzerman2004}.

\begin{figure*}
\begin{center}
\includegraphics[width=1.0\textwidth]{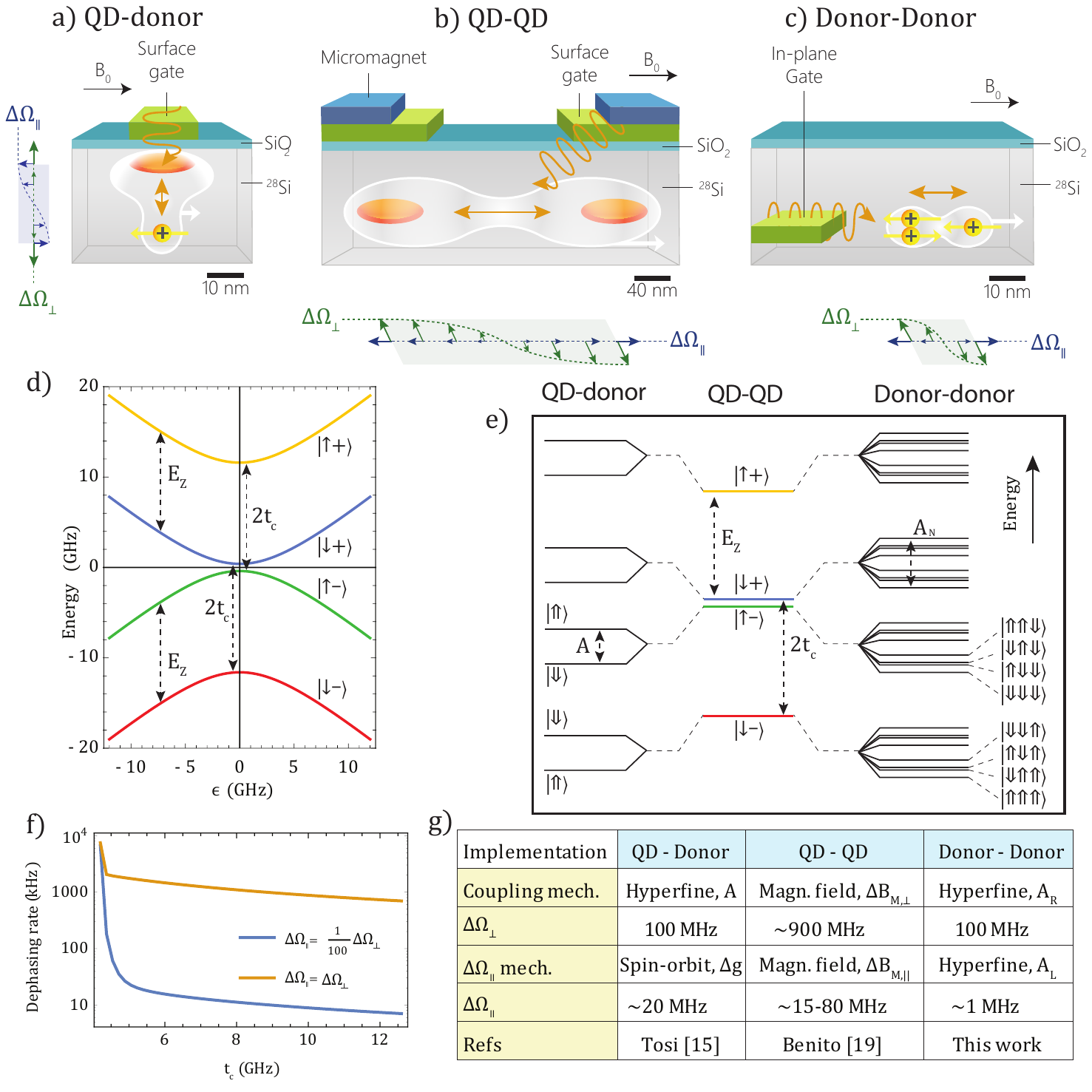}
\end{center}
\vspace{-0.5cm}
\caption{{\bf Flopping-mode electric-dipole spin resonance qubits and their properties.} Three different flopping-mode EDSR qubits implemented using a) quantum dot-donor, b) quantum dot-quantum dot, and c) donor-donor sites. The longitudinal (blue) and transverse (green) magnetic field gradients, $\Delta\Omega_{\parallel}$ and $\Delta\Omega_{\perp}$ are shown next to the different implementations. The quantum dot-donor and donor-donor implementations both use the hyperfine interaction from the electron-nuclear spins that are naturally present in donor systems to generate a spin-orbit coupling. The quantum dot-quantum dot system requires an additional micromagnet to create a spatially-varying magnetic field to induce an artificial spin-orbit coupling. The electron wavefunction is shown as the white cloud with a spin orientated parallel to the external magnetic field, $B_0$. The donor nuclei are shown as yellow positive charges. d) The energy spectrum for a single electron in a magnetic field ($E_z = \gamma_e B_0$) near the charge degeneracy between two different charge states with tunnel coupling, $t_c$. The energy spectrum at $\epsilon = 0$ for e) quantum dot-donor, quantum dot-quantum dot, and donor-donor implementations show the additional nuclear spin states for donor systems. f) The qubit dephasing rate for different longitudinal magnetic field gradients, $\Delta\Omega_{\parallel} = \Delta\Omega_\perp$ (yellow) and $\Delta\Omega_{\parallel} =\Delta\Omega_\perp/100$ (blue) with $\Delta\Omega_\perp=117\,\text{MHz}$. The smaller the longitudinal magnetic field gradient the more gradual the change in qubit energy, which results in lower errors over a larger detuning range. g) Summary of the effective magnetic field gradients found in the different flopping-mode EDSR qubits.}
\vspace{-0.5cm}
\label{fig:1}
\end{figure*}

In Fig.~\ref{fig:1}a)-c) we describe three different flopping-mode qubits in silicon. The two magnetic field gradients, $\Delta\Omega_{\perp}$ (single-qubit gate speed) and $\Delta\Omega_{\parallel}$ (qubit dephasing) present in each design arises from different physical mechanisms. Figure~\ref{fig:1}a) shows the quantum dot-donor hybrid qubit (flip-flop qubit)~\cite{Tosi2017}. Here, the spin-charge coupling arises from the hyperfine interaction of the electron spin with the nuclear spin of a single phosphorus donor which can be used to generate electron-nuclear spin quantum dot-donor transitions~\cite{PhysRevLett.102.027601}. The flopping-mode operation EDSR is performed by positioning the electron in a superposition of charge states between the donor nuclei and an interface quantum dot created using electrostatic gates. In this charge superposition state the hyperfine interaction is known to change from $A \approx 117$ MHz on the donor to $A \approx 0$ MHz on the quantum dot~\cite{Tosi2017}. The qubit states are $\ket{0} \equiv \ket{\Uparrow \downarrow}$ and $\ket{1} \equiv \ket{\Downarrow \uparrow}$ ($\ket{\textnormal{nuclear spin, electron spin}}$). The transverse magnetic field gradient, $\Delta\Omega_{\perp}$ (green) in this case arises from the changing hyperfine interaction as the electron is moved away from the donor nucleus. This voltage dependent hyperfine can then be used to resonantly drive the qubit states by applying an oscillating electric field. The longitudinal magnetic field gradient, $\Delta\Omega_{\parallel}$ (blue) is created by the difference in the electron $g$-factor between the quantum dot and donor such that the qubit energy differs whether the electron resides on the quantum dot or the donor.

The second flopping-mode qubit implementation, shown in Fig.~\ref{fig:1}b) is the quantum dot-quantum dot system~\cite{PhysRevB.100.125430}. Here the qubit states are the pure spin states of the electron in the ground charge state of the double quantum dot system, $\ket{0} \equiv \ket{\downarrow}$ or $\ket{1} \equiv \ket{\uparrow}$. The transverse magnetic field gradient, $\Delta\Omega_{\perp}$ required to drive qubit rotations is generated by an additional  micromagnet ($\sim 300$ nm away) designed to create a large magnetic field gradient ($\sim 10$ mT) across the quantum dots~\cite{Jang2020}. The flopping-mode EDSR is performed by biasing a single electron to a superposition between two charge states of different quantum dots and applying an oscillating electric field on resonance with the qubit energy. The stray field of the micromagnet is known to create a magnetic field gradient parallel to the external magnetic field corresponding to $\Delta\Omega_{\parallel}$ which leads to dephasing of the qubit.  

In this paper we propose an asymmetric donor quantum dot flopping-mode qubit shown in Fig.~\ref{fig:1}c). In this implementation the qubit utilises the hyperfine interaction to create a flip-flop transition of an electron spin with a nuclear spin on only one of the quantum dots. The other nuclear spins on the second quantum dot are then used a resource to reduce the dephasing rate of the qubit. This builds on a previous proposal where the electron spin could be electrically controlled by simultaneously flip-flopping with all nuclear spins across both quantum dots~\cite{wang2017edsr}. In principle, each donor quantum dot can be defined by any number of nuclear spins. Whilst a 1P-1P configuration is possible~\cite{Osika2021}, here we consider an asymmetric donor system to reduce the dephasing anticipated from the longitudinal magnetic field gradient. As the number of donors comprising the quantum dot is increased, the hyperfine strength of the first electron on that quantum dot becomes larger~\cite{Wang2016a}. This is useful for increasing the transverse magnetic field gradient required for qubit driving and can make the hyperfine interaction significantly different between the quantum dots to selectively drive particular flip-flop transitions. However, the larger hyperfine on the secondary quantum dot also makes the longitudinal magnetic field gradient larger. To reduce this effect, we propose filling one of the quantum dots with more electrons to create a shielding effect of the outer electron to the donor nuclear spins. This results in a reduced hyperfine coupling~\cite{Wang2016a} and lower dephasing rate for any orientation of nuclear spins. In particular, we consider the specific case of a single donor coupled to a 2P quantum dot (2P-1P) at the (2,1)$\leftrightarrow$(3,0) charge transition so that the two inner electrons on the 2P (left) quantum dot lower the hyperfine interaction of the outermost electron. Nuclear spin control of the donors in the 2P quantum dot allows further engineering of the total hyperfine coupling experienced by the electron. As we will show later, this reduces the longitudinal magnetic field gradient, $\Delta \Omega_{\parallel}$ and leads to increased coherence times. The qubit states are $\ket{0} \approx \ket{\Downarrow\Uparrow \Uparrow \downarrow}$ and $\ket{1} \approx \ket{\Downarrow\Uparrow \Downarrow \uparrow}$ which are coupled via a flip-flop transition of the electron with the 1P (right) nuclear spin. Such a donor-donor implementation therefore also uses the hyperfine interaction from the electron-nuclear spin system to drive qubit transitions as with the flip-flop qubit in Fig.~\ref{fig:1}a). The key difference is that the magnetic field gradient can be engineered during fabrication by controlling the number of donors in each quantum dot. Since the hyperfine interaction is known to change considerably for multi-donor quantum dots we can make $\Delta\Omega_{\perp}$ up to $\sim 300$ MHz and $\Delta\Omega_{\parallel}$ less than a few MHz~\cite{Hileeaaq1459}, see Fig.~\ref{fig:1}g). This is in contrast to the flip-flop qubit where $\Delta\Omega_{\parallel}$ is determined by the difference in the electron $g$-factor on the donor atom and the quantum dot, the latter being known to vary due to atomic steps at the interface where the quantum dot is formed~\cite{PhysRevB.97.241401}.

In this paper we will show that the additional nuclei in these multi-donor quantum dots can be used to minimise the dephasing rate of the qubit. This is because the strength of the hyperfine interaction with the nuclear spins that are not flipping with the electron spin largely determines the dephasing rate. By engineering the hyperfine strength on the multi-donor quantum dots, we therefore maximise the coherence time of the EDSR qubit. By directly controlling the nuclear spin states and the number of electrons on the double donor flopping-mode EDSR qubit, we can also operate over a wide range of magnetic fields and tunnel couplings. Most importantly, the qubit shows low errors, $ < 10^{-3}$, below the error threshold for surface code error correction, with realistic noise levels in isotopically purified silicon-28~\cite{Muhonen2014,Kranz2020}. The robustness of the qubit to magnetic field and tunnel coupling variations is particularly useful for scaling to large qubit arrays where inevitable imperfections in fabrication can reduce qubit quality. Finally, we show that the low error rate and the spin-charge coupling predicted for the qubit will allow for strong-coupling to superconducting microwave cavities. This spin-cavity coupling has been systematically studied by Osika et al.~\cite{Osika2021} who consider the specific case of a 1P-1P double donor system. They show that the use of a symmetric hyperfine coupling in a 1P-1P or the recently discovered electrically induced spin-orbit coupling~\cite{Weber2018}, allows for strong-coupling of a phosphorus-doped silicon qubit to a superconducting cavity (simulated using finite element modelling). These two papers highlight multiple routes for achieving two-qubit couplings between Si:P qubits via superconducting microwave resonators.

A generic energy level spectrum for all flopping-mode EDSR qubits is shown in Fig.~\ref{fig:1}d). The spectrum describes a single electron near the degeneracy point of two different charge states as a function of the detuning between them, $\epsilon$ (that is at $\epsilon = 0$ the charge states are equal in energy). The charge states have a tunnel coupling, $t_c$ and the electron spin states are split by the Zeeman interaction, $E_z = \gamma_e B_0$ in a static magnetic field, $B_0$, where $\gamma_e$ is the electron gyromagnetic ratio. The system is described by the spin of the single electron and the bonding/anti-bonding charge states ($\ket{+} = (\ket{L} + \ket{R})/\sqrt{2}$ and $\ket{-} = (\ket{L} - \ket{R})/\sqrt{2}$ where $\ket{L}$ and $\ket{R}$ are the left and right quantum dot orbitals, respectively) resulting in a set of four basis states $\{\ket{\downarrow -}, \ket{\uparrow -}, \ket{\downarrow +}, \ket{\uparrow +} \}$ corresponding to the red, green, blue, and yellow states in Fig.~\ref{fig:1}d). The spin-charge coupling is maximised when the charge ground state $\ket{\uparrow -}$ (green) hybridises with the charge excited state $\ket{\downarrow +}$ (blue), which at $\epsilon = 0$ occurs when $E_z \approx 2 t_c$ (see Fig.~\ref{fig:1}d)). In donor-based systems these electron spin states are split due to the hyperfine interaction of the electron with the quantised nuclear spin states. In Fig.~\ref{fig:1}e) we show a comparison of the energy levels involved for the quantum dot-donor, quantum dot-quantum dot and donor-donor implementations at $\epsilon = 0$. The quantum dot-quantum dot system is comprised of only charge and electron spin states. The presence of nuclear spins in donor systems increases the number of states by a factor of $2^n$ where $n$ is the number donors (we note that the donor-quantum dot flopping mode qubit has 8 combined electron, nuclear and charge states and our proposal for a 2P-1P system has 32, see Fig.~\ref{fig:1}e)). Importantly, for operation of the donor-based EDSR qubit the electron and nuclear spins must be anti-parallel, $\ket{\Uparrow \downarrow}$ or $\ket{\Downarrow \uparrow}$ to allow for the flip-flop transition. Whilst the qubit does not need a micromagnet to generate a spin-charge coupling, it is important to minimise any unwanted nuclear spin flip-flop transitions which can lead to leakage out of the computational basis. We will show that the added leakage pathways from the nuclear spins can be largely controlled by Gaussian pulse shaping, leading to error rates on the order of $10^{-4}$. In the long term this can be improved further using pulse shaping techniques such as derivative reduction by adiabatic gates (DRAG~\cite{PhysRevLett.103.110501} which we do not consider in this work).

\begin{figure*}
\begin{center}
\includegraphics[width=1.0\textwidth]{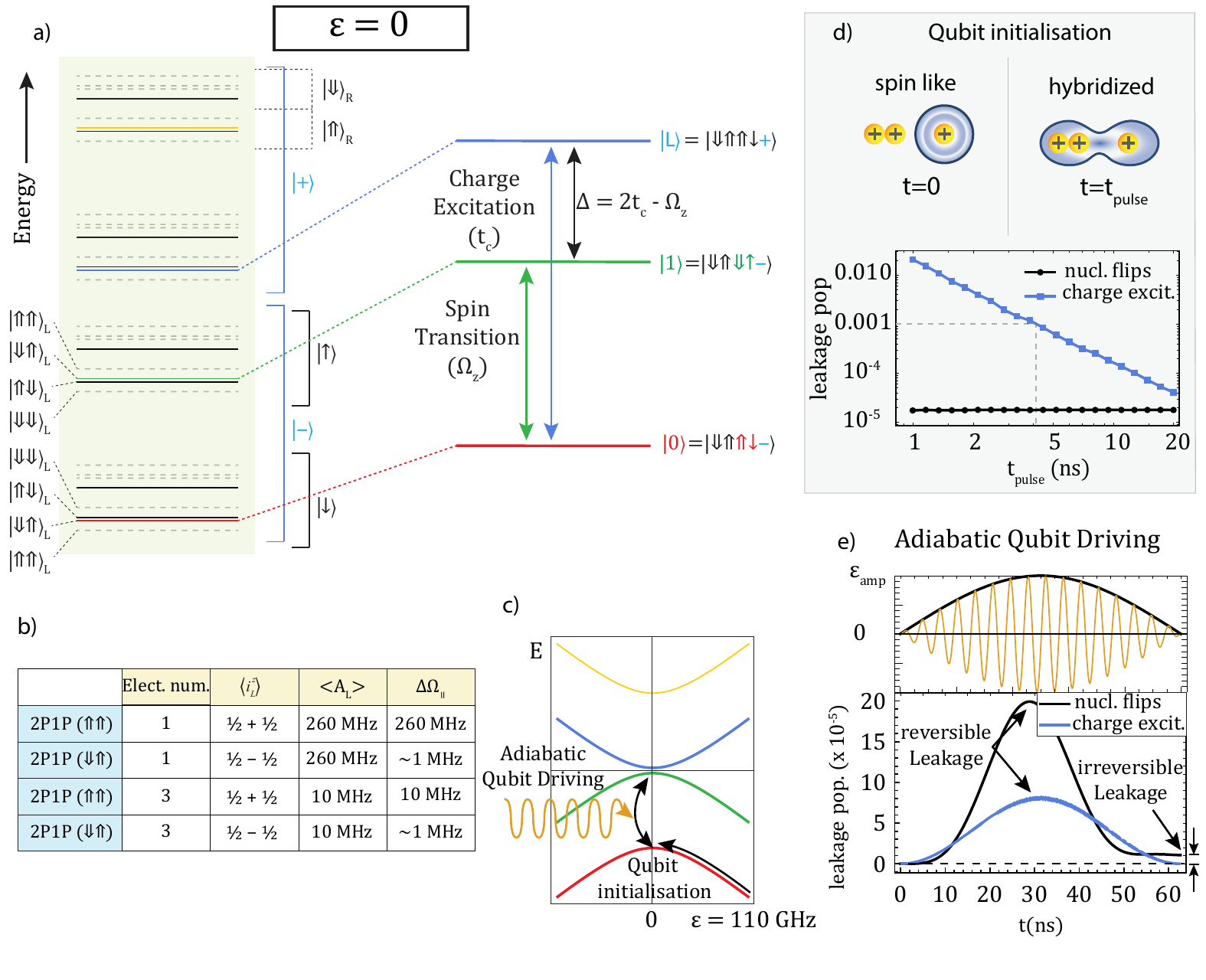}
\end{center}
\vspace{-0.5cm}
\caption{{\bf Operation of the donor-donor flopping mode qubit.} Due to spin conservation, only a subset of the nuclear spin states in the hyperfine manifold in {\bf a)} need to be considered for qubit operation. For a 2P-1P donor-donor device, the qubit states are displayed in red and green, the lowest (highest) excited charge state in blue (yellow), the nuclear spin leakage states where the total spin of the system is conserved are shown in black. The leakage probability of the nuclear spin states can be minimised by careful pulse design. The states not involved in the qubit operation (other nuclear spin states with no leakage pathway) are shown as dashed grey lines. {\bf b)} Control of the electron number using electrostatic gates and nuclear spin orientation ($\left<i^z_{L}\right>$) using NMR allows us to tune the hyperfine coupling, $\left< A_L \right>$ and longitudinal magnetic field gradient $\Delta\Omega_{\parallel}$. {\bf c)} Leakage out of the qubit subspace needs to be considered both when initialising the qubit for control and when driving the qubit at $\epsilon=0$. {\bf d)} Initialisation of the qubit ground state for a 2P-1P donor-donor qubit at the (3,0) $\leftrightarrow$ (2,1) electron configuration from the localised electron state (at $\epsilon=110\,\text{GHz}$) to the hybridized state (at $\epsilon=0$), using a variable pulse time $t_\text{pulse}$, at $B=0.3\,\text{T}$, $t_c=5.9\,\text{GHz}$. The qubit population that leaks into the excited charge states and other nuclear spin states at the end  of the transfer are displayed as a function of the pulse time. {\bf e)} Driving of the qubit states using microwave pulses allows full control of the qubit states. Gaussian pulse shaping allows for the reversal of state leakage during the qubit operation (top). We show the charge (blue) and nuclear spin (black) leakage probabilities during the $\pi/2 - X$ Gaussian pulse for the donor-donor qubit using optimal parameters for this device, drive amplitude of $\epsilon_\text{amp}=0.9\,\text{GHz}$ at $B=0.23\,\text{T}$, and $t_c=5.6\,\text{GHz}$ (bottom). The irreversible leakage for the the nuclear spin states is $\sim1 \times 10^{-5}$ well below the 1\% error required for fault tolerance.}
\vspace{-0.5cm}
\label{fig:2}
\end{figure*}

Minimising the magnetic field gradient $\Delta\Omega_{\parallel}$ parallel to $B_0$ is important to prevent dephasing of the qubit. The longitudinal magnetic field gradient arises from either the stray field of the micromagnet~\cite{Kha2015, Struck2020} or from the isotropic hyperfine interaction~\cite{Wang2016a}, that takes the form $A(s_x i_x + s_y i_y + s_z i_z)$ in the Hamiltonian, where $s_i$ ($i_i$) is the electron (nuclear) spin operator. The fact that the hyperfine interaction is isotropic means that irrespective of the magnetic field orientation there will always be some hyperfine component parallel to the external magnetic field resulting in an energy gradient $\Delta\Omega_{\parallel}$ (with respect to detuning). Since charge noise couples to the qubit via charge detuning, the smaller this gradient, the flatter the qubit energy as a function of detuning, and the lower the charge noise induced dephasing during qubit operation. In Fig.~\ref{fig:1}f) we plot the qubit dephasing rate as a function of tunnel coupling at $\epsilon = 0$ (where the qubit drive is performed) for two different values of $\Delta\Omega_{\parallel} = \Delta\Omega_{\perp}/100$ MHz (small $\Delta\Omega_{\parallel}$) and $\Delta\Omega_{\parallel} = \Delta\Omega_{\perp}$ MHz (large $\Delta\Omega_{\parallel}$). We can see that the qubit dephasing rate remains smaller over a wider range of tunnel couplings for small $\Delta\Omega_{\parallel}$ compared to large $\Delta\Omega_{\parallel}$ indicating that the qubit will perform better when $\Delta\Omega_{\parallel}$ is minimised. In general, for flopping-mode qubits it is beneficial to maximise $\Delta\Omega_{\perp}$ (qubit driving) and to minimise $\Delta\Omega_{\parallel}$ (qubit dephasing). To summarise, in Fig.~\ref{fig:1}g) we compare the physical parameters that would be expected for the three different flopping-mode EDSR qubit implementations. We can see that the quantum dot-quantum dot implementation obtains very large $\Delta\Omega_{\perp} \sim 900$ MHz allowing for fast qubit operations; however, $\Delta\Omega_{\parallel} \sim 15-80$ MHz is also relatively high leading to faster qubit dephasing. The quantum dot-donor and donor-donor qubits both have similar $\Delta\Omega_{\perp} \sim 100$ MHz values due to similar hyperfine interaction strengths from the phosphorus donor. However, by minimising the hyperfine interaction on the multi-donor quantum dot instead of the difference in $g$-factors, we can achieve $\Delta\Omega_{\parallel} \sim 0$ MHz for the donor-donor EDSR qubit, smaller than other flopping mode qubits. At the same time the donor-donor implementation operates away from interfaces that lead to charge noise and do not require additional micromagnets which can also induce charge noise~\cite{Kha2015}. In the next sections we theoretically investigate the fidelity of single-qubit gates and microwave cavity coupling for two-qubit gates. In particular, we focus on the benefits of using two different size donor quantum dots (2P-1P) for flopping-mode EDSR to maximise $\Delta\Omega_{\perp}$ and minimise $\Delta\Omega_{\parallel}$ by controlling the nuclear spins and the electron shell filling on both donor-based quantum dots.

The qubit we propose utilises flopping-mode EDSR to electrically drive the electron-nuclear flip-flop transition where the two charge sites are defined by donor-based quantum dots. The Hamiltonian for a single electron between two tunnel coupled donor-based quantum dots approximately 10 - 15 nm apart with $N_L$ (donors in the left quantum dot) and $N_R$ (donors in the right quantum dot) is given by,
\begin{equation}
H = H_{Zeeman} + H_{Charge} + H_{Hyperfine},
\label{eq:Ham}
\end{equation}
where $H_{Zeeman} = \gamma_e B_0 s_z + \gamma_n B_0 \sum i_z$ is the Zeeman term for both the electron ($\gamma_e \approx 27.97$ GHz, the electron gyromagnetic ratio) and nuclear spins ($\gamma_n \approx -17.41$ MHz, the nuclear gyromagnetic ratio), $H_{Charge}$ describes the tunnel coupling, $t_c$ and detuning, $\epsilon$ between the charge states of the donors that have an excess electron on one of the quantum dots ($2n_l, 2n_r+1$) $\leftrightarrow$ ($2n_l+1, 2n_r$) and $H_{Hyperfine}$ represents the detuning dependent contact hyperfine interaction ($A_L$ and $A_R$ for the left and right quantum dots) of the outermost electron spin to each of the $N_L + N_R$ phosphorus nuclear spins (see Appendix A).

In principle, each quantum dot can be formed by any number of phosphorus donors; however, here we investigate the specific case of $N_L = 2$ and $N_R = 1$, that is, the 2P-1P system (see Fig.~\ref{fig:2}a) for the energy level diagram at $\epsilon = 0$). The qubit states are defined as $\ket{0} \approx \ket{\Downarrow\Uparrow \Uparrow \downarrow -}$ and $\ket{1} \approx \ket{\Downarrow\Uparrow \Downarrow \uparrow -}$ and a transition between the two states corresponds to a flip-flop of the electron spin with the nuclear spin on the right donor quantum dot. The nuclear spin states on the left donor quantum dot remain unchanged during the transition. The charge state $\ket{-}$  is defined by the two quantum dot orbitals associated with the ($3,0$) $\leftrightarrow$ ($2, 1$)  charge transition. To compare the donor-donor flopping-mode qubit to the quantum dot-quantum dot and quantum dot-donor implementations we approximate the Hamiltonian in Eq.~\ref{eq:Ham} using a Schrieffer-Wolff transformation to a general flopping-mode Hamiltonian in terms of the transverse ($\Delta \Omega_{\perp}$) and longitudinal ($\Delta\Omega_{\parallel}$) gradients (see Appendix A),
\begin{equation}
H = \frac{\Omega_z}{2} \sigma_z + \epsilon \tau_z + t_c \tau_x + \Big( \frac{\Delta\Omega_{\parallel}}{4} \sigma_z + \frac{\Delta\Omega_{\perp}}{4} \sigma_x \Big) \tau_z.
\label{eq:genHam}
\end{equation}
Equation~\ref{eq:genHam} is written in a similar format to Eq.~\ref{eq:Ham} where $\sigma_i$ ($\tau_i$) are the Pauli-operators for the combined electron-nuclear spin (charge) degree-of-freedom. The first term, $\Omega_z$ is the energy of the combined electron-nuclear spin state (which depends on the exact value of the left and right donor hyperfine, $A_L$ and $A_R$),
\begin{equation}
\Omega_z = \sqrt{\Omega_s^2 + A_R^2/4},
\end{equation}
where $\Omega_s = (\gamma_e + \gamma_n)B_0 + \sum_k^{N_L} A_{L,k} \left<i^z_{L,k}\right>/2$ is the Zeeman energy with a correction due to the hyperfine interaction of the electron with the nuclear spins in the left quantum dot and $\left<i^z_{L,k}\right>$ is the expectation value of the $z$-projection of the $k$-th nuclear spin on the left quantum dot. The charge part of the Hamiltonian is described by the second (detuning, $\epsilon$) and third (tunnel coupling, $t_c$) terms of Eq.~\ref{eq:genHam}. The last term in Eq.~\ref{eq:genHam} corresponds to the charge-dependent hyperfine interaction,
\begin{equation}
\Delta\Omega_{\parallel} = \sum_k^{N_L} A_{L,k} \left<i^z_{L,k}\right> \cos{\theta} - A_R \sin{\theta},
\end{equation}
\begin{equation}
\Delta\Omega_{\perp} =  A_R \cos{\theta} - \sum_k^{N_L} A_{L,k} \left<i^z_{L,k}\right> \sin{\theta},
\end{equation}
where $\tan{\theta} = A_R/(2\Omega_s)$. Since $\Omega_s$ is typically $> 5$ GHz is generally much greater than $A_R \approx 100$ MHz, $\sin{\theta} \approx 0$ and $\cos{\theta} \approx 1$ then $\Delta\Omega_{\parallel} \approx \sum_k^{N_L} A_{L,k} \left<i^z_{L,k}\right>$ and $\Delta\Omega_{\perp} \approx A_R$. This means that we can control $\Delta\Omega_{\parallel}$ during fabrication by engineering the number of the donor atoms in each quantum dot. Additionally, during qubit operation we can optimise $\Delta\Omega_{\parallel}$ by controlling the nuclear spins on the left quantum dot using nuclear magnetic resonance (NMR)~\cite{Pla2013} or dynamic nuclear polarisation~\cite{PhysRevLett.100.067601}, and by controlling the electron shell filling in the left quantum dot. Figure~\ref{fig:2}b) shows a table of different nuclear spin and electron configurations determining the magnitude of the hyperfine coupling strengths $A_{L,k}$ and their effect on the value of $\Delta\Omega_{\parallel}$. In general, the larger the quantum dot the larger $\sum A_{L,k}$ since the phosphorus donors create a stronger confinement potential for the electron which increases the contact hyperfine strength. However, by adding a pair of electron spins to the left quantum dot (increasing the total electron number from 1 to 3), the two innermost electrons form an inactive singlet-state that screens the outermost electron defining the qubit from the nuclear potential of the donors. The shielding decreases $\sum A_{L,k}$ and results in longer dephasing times. Furthermore, the presence of more then one donor in the left quantum dot allows a further reduction of the longitudinal gradient $\Delta\Omega_\parallel$ by controlling their nuclear spins. From Fig.~\ref{fig:2}b) we can see that by using antiparallel nuclear spin states ($\left<i^z_{L,1}\right> = 1/2$ and $\left<i^z_{L,2}\right> = -1/2$) on a 2P quantum dot we can lower the value of $\Delta\Omega_{\parallel}$ to close to 0. This ability to control the number of electrons and nuclear spin states on the left quantum dot forms the motivation for operating the qubit using $\ket{0} \approx \ket{\Downarrow\Uparrow \Uparrow \downarrow -}$ and $\ket{1} \approx \ket{\Downarrow\Uparrow \Downarrow \uparrow -}$ at the ($3,0$) $\leftrightarrow$ ($2, 1$) transition. Note that the nuclear spin states $\ket{\Uparrow\Downarrow}$ and $\ket{\Downarrow\Downarrow}$ for the 2P are equivalent to $\ket{\Downarrow\Uparrow}$ and $\ket{\Uparrow\Uparrow}$, respectively and so were not explicitly included in Fig.~\ref{fig:2}b).

Additional nuclear spin states could create more leakage pathways out of the computational basis, but here we show that these additional nuclear spins behave as a resource and are not a limiting factor for the qubit operation. In particular, there are two crucial steps in the qubit operation where leakage from the computational basis can occur: during initialisation and during driving of single-qubit gates.

\begin{figure}
\begin{center}
\includegraphics[width=1\columnwidth]{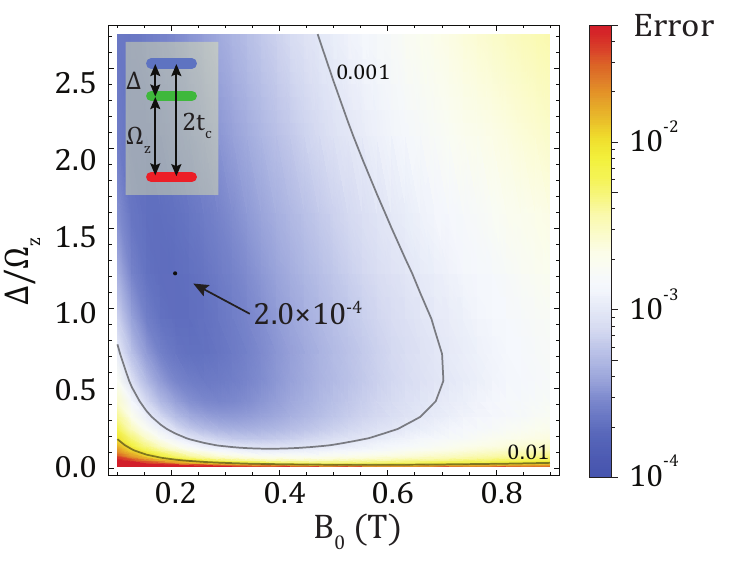}
\end{center}
\vspace{-0.5cm}
\caption{{\bf $\pi/2$-gate error of the all-epitaxial flopping-mode EDSR donor-based qubit.} Qubit error with $\Delta A_L = |A_{L,1} - A_{L,2}| = 1$ MHz as a function of the external magnetic fields $B_0$ and spin-charge detuning $\frac{\Delta}{\Omega_z} = \frac{2t_c - \Omega_z}{\Omega_z}$ at $\epsilon = 0$. The gate error remains below $10^{-3}$ over a magnetic field range of 0.4 T and for $\Delta/\Omega_z$ values from 0.5 to more than 2.5. The optimal operating point with a minimum error of $2\times10^{-4}$ is shown at the black dot. The inset shows the 3-level energy diagram for the qubit with energy, $\Omega_z$, tunnel coupling, $t_c$ and spin-charge detuning, $\Delta = 2t_c - \Omega_z$.}
\vspace{-0.5cm}
\label{fig:3}
\end{figure}

First, we will describe and examine the initialisation process for potential charge and nuclear spin state leakage. Excited charge state leakage is present in all flopping-mode EDSR based qubits due to the hybridisation of charge and spin. For $|\epsilon| \gg t_c$ there is no charge-like component of the qubit and the ground state can be initialised simply by loading a $\ket{\downarrow}$ electron from a nearby electron reservoir~\cite{Elzerman2004}. The nuclear spins can also be initialised via NMR~\cite{Pla2013} or dynamic nuclear polarisation~\cite{PhysRevLett.104.226807} to place the nuclear spin in the $\ket{\Uparrow}$ state. Next, the detuning is ramped to $\epsilon = 0$ to initialise the $\ket{0}$ qubit state, see Fig.~\ref{fig:2}c). During the ramp the qubit can leak out of the computational basis via charge excitation into the excited charge state or through unwanted nuclear spin flips (see Appendix B). In Fig.~\ref{fig:2}d) we show the simulated leakage probability of a donor-based flopping mode qubit for both of these leakage pathways during the initialisation ramp as a function of ramp time with $t_c = 5.6$ GHz, $\Delta A_L = |A_{L,1} - A_{L,2}| = 1$ MHz and $B_0 = 0.23$ T. We can see that regardless of the initialisation pulse time, $t_{pulse}$ the leakage into the excited charge states (blue line in Fig.~\ref{fig:2}d)) is the dominant pathway compared to the nuclear spin leakage (black line in Fig.~\ref{fig:2}d)). The nuclear spin leakage is much lower compared to the charge leakage because the probability of a flip-flop transition away from $\epsilon = 0$ is small since the hyperfine strength changes very slowly with detuning compared to the charge states and the nuclear spin leakage states are weakly coupled to the qubit states. The charge leakage mechanism exists for all flopping-mode EDSR based qubits due to the non-adiabaticity of the initialisation pulse. By ramping slow enough however, we can initialise the qubit at $\epsilon = 0$ with a leakage error of $10^{-3}$ for a $t_{pulse} = 4$ ns ramp. The nuclear spin leakage does not depend heavily on the pulse time and remains well below the charge leakage with an error of $\sim 2 \times 10^{-5}$. Therefore, we can conclude that the nuclear spin state leakage is not a limiting factor in the initialisation of the qubit.

In Fig.~\ref{fig:2} a) we show the full energy spectrum of the donor-donor implementation at zero detuning, $\epsilon = 0$. On the right we show the qubit states (red and green) and the lowest charge leakage state (blue) with their relative energies. There are 32 spin and charge states in the full system (black and grey). Two types of leakage errors can occur during driving due to the presence of the nuclear spin states in the 2P-1P donor-based flopping-mode qubits (see Appendix C for a detailed discussion). These two leakage errors only become critical for nearly degenerate nuclear spin states. This can be the case when the hyperfine values are similar, for example when $A_{L,k} \approx A_R$. The first leakage error in the 2P-1P donor-based flopping-mode qubits is due to an unwanted electron-nuclear flip-flop transitions with the nuclear spins in the left quantum dot such as the transition $\ket{\Downarrow\Uparrow \Uparrow \downarrow -} \rightarrow \ket{\Downarrow\Downarrow \Uparrow \uparrow -}$ and is proportional to $(A_{L,k}/A_R)^2$. Therefore, it is optimal to make $A_{L,k} \ll A_R$ to limit the unwanted flip-flop events. This is easily achieved by creating asymmetric donor-based quantum dots since the hyperfine strength depends on the number of donors and the presence of inactive electron shells in the quantum dot~\cite{Wang2016a}. The second leakage process involves an unlikely simultaneous electron-nuclear flip-flop with all three of the nuclear spins (for example, $\ket{\Downarrow\Uparrow \Uparrow \downarrow -} \rightarrow \ket{\Uparrow\Downarrow \Downarrow \uparrow -}$). For the corresponding error to be small, the energy gap $\Delta A_L/4$ between the qubit states and the nearest leakage state needs to be non zero. This is likely the case due to the presence of electric fields in a real device and so this leakage pathway is easily avoidable. Leakage states have been extensively investigated in the superconducting qubit community~\cite{Theis_2018}. Well designed pulses have minimised leakage out of the computational basis by adiabatically reversing the leakage process~\cite{PhysRevLett.103.110501}. The simulations performed for the remainder of the paper use a Gaussian pulse shape~\cite{PhysRevA.82.042339} (shown in Fig.~\ref{fig:2}e) top) to partially reverse the leakage process due to charge and nuclear spins. Using a Gaussian pulse does not fully reverse the leakage process and inevitably there will be some leakage error at the end of qubit gate, see Fig.~\ref{fig:2}e) bottom at the end of the pulse ($t \approx 65$ ns).

\begin{figure}
\begin{center}
\includegraphics[width=1\columnwidth]{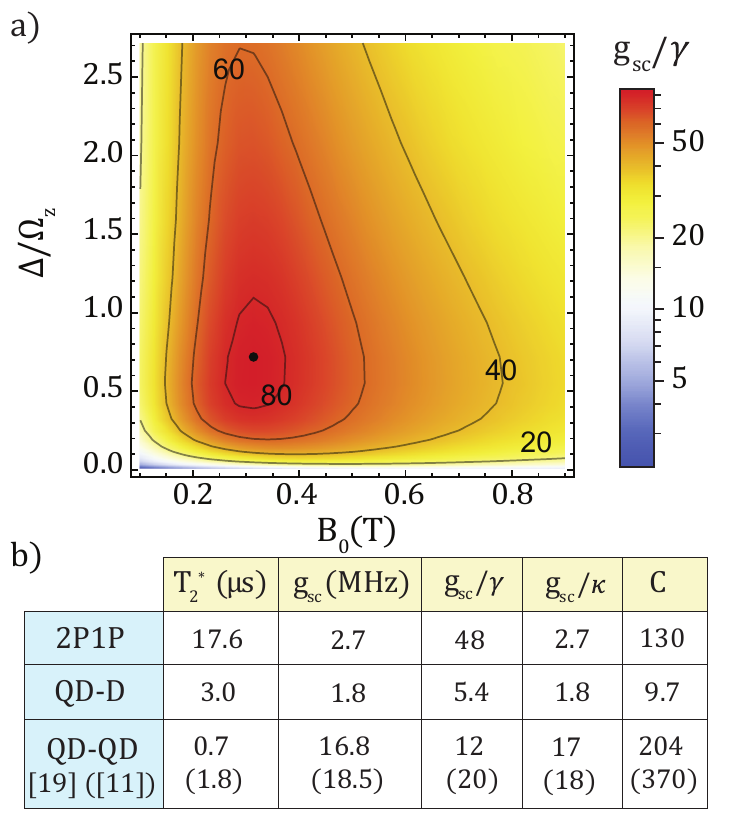}
\end{center}
\vspace{-0.5cm}
\caption{{\bf  Strong coupling of the all-epitaxial flopping mode qubit to a superconducting cavity resonators.} {\bf a)} For the 2P-1P with the 2P nuclear spins in $\ket{\Downarrow\Uparrow}$ at the (2,1) $\leftrightarrow$ (3,0) charge transition, ratio of the spin-cavity coupling strength, $g_{sc}$ to the qubit decoherence rate, $\gamma$ as a function of the spin-charge relative detuning $\Delta/E_\text{Z}$ and the external magnetic field, $B_0$. We assume charge coupling of the qubit to cavity to be 100 MHz. {\bf b)} Table of the main qubit-cavity coupling characteristic values for different flopping mode implementations. The cooperativity is defined as the product of $g_{sc}/\gamma$ and $g_{sc}/\kappa$. For each implementation, all value are calculated at the tunnel coupling and magnetic field value where $C$ is a maximum under the condition that the qubit drive error is below 0.1\% (not necessarily where $g_{sc}/\gamma$ is the largest) and is therefore lower than the maximum achievable coupling of $g_{sc}/\gamma = 85$ in a). For the QD-D qubit~\cite{Tosi2017}, we chose $\Delta\Omega_\parallel=117\,\text{MHz}$, and  $\Delta\gamma=-0.2\%$ corresponding to $\Delta\Omega_\parallel=11\,\text{MHz}$ at $B=0.2\,\text{T}$. For the QD-QD qubit we chose gradient values as cited in \cite{PhysRevB.100.125430} and \cite{PhysRevResearch.2.012006} resp. In Benito et.al~(\cite{PhysRevB.100.125430}), $\Delta\Omega_\perp=0.96\,\text{GHz}$ (corresponding to $2b_x=4\,\mu\text{eV}$) and $\Delta\Omega_\parallel=78\,\text{MHz}$ (corresponding to $2b_z=0.32\,\mu\text{eV}$). In Croot et. al~(\cite{PhysRevResearch.2.012006}), $\Delta\Omega_\perp=0.84\,\text{GHz}$ (corresponding to $2b_x=30\,\text{mT}$) and $\Delta\Omega_\parallel=15\,\text{MHz}$ (corresponding to $2b_z\approx0.5\,\text{mT}$) }
\vspace{-0.5cm}
\label{fig:4}
\end{figure}

To further investigate the qubit performance in Fig.~\ref{fig:3} we show the qubit error for a $\pi/2-X$ gate as a function of magnetic field and tunnel coupling including dephasing, relaxation and leakage errors (see Appendix C). Importantly, the gate error remains low ($< 10^{-3}$) over a wide range of magnetic fields ($\sim 0.1 - 0.5$ T) and  for relative changes in the tunnel couplings of more then 300\%, corresponding to a tolerance of more then $8\, (17)\,\text{GHz}$ at $B=0.2\,(0.4)\,\text{T}$. We note that the other flopping-mode qubits have only been optimised over a much smaller parameter space, confined to the location of so-called error sweetspots, that restrict the operational range of magnetic field and tunnelcouplings~\cite{Tosi2017,PhysRevB.100.125430}. The wide operational parameter space is crucial in a large-scale architecture with a fixed magnetic field where small uncertainties in the tunnel coupling during fabrication can lead to variation in the qubit performance. The large range of tunnel couplings where the donor-donor qubit can operated means that these small uncertainties will not be detrimental to the overall quantum computer performance. By optimising the magnetic field and tunnel coupling during fabrication we can achieve a minimum gate error of $2.0 \times 10^{-4}$ well below the surface code fault-tolerant threshold.

Finally, we examine the suitability of the proposed flopping-mode qubit for two-qubit couplings. Due to the charge-like character, the flopping-mode qubit can be coupled directly via the charge dipole interaction~\cite{Tosi2017}. The range of the dipole interaction can be extended using floating gate structures~\cite{PhysRevX.2.011006} or by coupling two qubits to a superconducting microwave resonator~\cite{Viennot408}. Indeed, one of the most attractive properties of spin-charge coupling is that it allows for coupling of single spins to microwave cavities which can be used for two-qubit gates between distant qubits~\cite{Mi156,Samkharadze1123}. Spin-cavity coupling is achieved by carefully designing the cavity frequency, $f_c$ to be on resonance with the qubit frequency, that is, $2 t_c \approx \gamma_e B_0 \approx f_c$. Recent high-kinetic inductance cavities have produced large zero-point voltage fluctuations on the order of a $20$ $\mu$V with photon loss rates on the order of $\kappa = 1$ MHz~\cite{PhysRevApplied.5.044004, Samkharadze1123}. For our donor-donor qubit this would correspond to a charge-cavity coupling on the order of tens of MHz. Following the detailed work in Osika~\emph{et al.}~\cite{Osika2021} where a specific implementation of the 1P-1P qubit is discussed we assume that the charge-cavity coupling is on the order of $100$ MHz. Note that the simulations in Osika~\emph{et al.}~\cite{Osika2021} were performed without the kinetic inductance of the superconductor and as such the $100$ MHz charge-cavity coupling should be taken as a lower bound.

In Fig.~\ref{fig:4}a) we plot the expected ratio of the spin-cavity coupling strength, $g_{sc}$ to the qubit dephasing rate, $\gamma$ for an optimised 2P-1P qubit with $\Delta\Omega_{\parallel} = 0.5$ MHz by initialising the nuclear spins in antiparallel states and using the 3 electron regime. The dephasing rate, $\gamma$ is calculated by converting the error probability into a coherence time based on the $\pi/2$ gate time for each value of $t_c$ and $B_0$ (see Appendix D). The qubit dephasing rate itself is smaller than $g_{sc}$ for all values of $t_c$ and $B_0$ shown indicating that qubit coherence is not the limiting factor in achieving the strong coupling regime. To achieve strong qubit-cavity coupling $g_{sc}$ also needs to be faster than the decay rate of the cavity such that the cooperativity is larger then one: $C = g_{sc}^2/\gamma \kappa > 1$. In Fig.~\ref{fig:4}b) we show the estimated coupling parameters for the different flopping-mode qubit implementations discussed in this work. Theoretical analysis of the EDSR protocol yields $T^*_2 = 17.6 \mu$s for the 2P-1P configuration. Taking this coherence time as a reasonable estimate of the spin dephasing rate for qubit-cavity coupling suggests that it would allow the strong-coupling limit to be reached, $g_sc/\gamma = 47.8$. The cooperativity of the 2P-1P qubit is comparable to the other flopping-mode EDSR systems, indicating that the qubit can also coupled to superconducting resonators for two-qubit gates. Note that all of the proposed implementations can reach the strong coupling regime with $C > 1$ allowing for two-qubit interactions using superconducting cavities.

In summary, we propose the implementation of a flopping-mode EDSR donor-based qubit and have performed detailed calculations of the error sources. The nuclear spins not directly involved in the qubit fip-flop transition can be used to engineer the longitudinal magnetic field gradient to increase the qubit coherence time. We show that the donor-donor molecule qubit can achieve error rates below the 1\% necessary for fault-tolerant quantum computation. The qubit can be operated over a wide range of magnetic field (0.4 T) and for relative variations in the tunnel coupling above 300\% ($\sim 5 - 20$ GHz). Fast, high-fidelity single-qubit gates with errors on the order of $10^{-4}$ are theoretically predicted, comparable to that found in other semiconductor qubits with full electrical control~\cite{Veldhorst2014,Yoneda2018}. Finally, we examined the possibility of coupling this qubit to a superconducting cavity resonator where we showed strong coupling is achievable with a cooperativity, $C \sim 130$. Based on the low qubit error rate, small qubit footprint, versatility in two-qubit coupling, and robustness to fabrication errors we have shown that flopping-mode EDSR based on two donor quantum dots provides an attractive route for scaling in donor-based silicon computing.

\section*{ACKNOWLEDGEMENTS}
The research was supported by the Australian Research Council Centre of Excellence for Quantum Computation and Communication Technology (project number CE170100012), the US Army Research Office under contract number W911NF?17?1?0202, and Silicon Quantum Computing Pty Ltd. M.Y.S. acknowledges an Australian Research Council Laureate Fellowship.

\bibliography{proposal_refs,felix_library,library}

\begin{thebibliography}{10}

\bibitem{Muhonen2014}
J.~T. Muhonen et~al.,
\newblock Nature Nanotechnology {\bf 9}, 986 (2014).

\bibitem{vahapoglu2021}
E.~Vahapoglu et~al.,
\newblock arXiv:2012.10225  (2021).

\bibitem{Buch2013}
H.~B{\"{u}}ch, S.~Mahapatra, R.~Rahman, A.~Morello, and M.~Y. Simmons,
\newblock Nature Communications {\bf 4}, 2017 (2013).

\bibitem{PhysRevB.78.195302}
E.~I. Rashba,
\newblock Phys. Rev. B {\bf 78}, 195302 (2008).

\bibitem{PhysRevB.67.115324}
L.~S. Levitov and E.~I. Rashba,
\newblock Phys. Rev. B {\bf 67}, 115324 (2003).

\bibitem{Nowack1430}
K.~C. Nowack, F.~H.~L. Koppens, Y.~V. Nazarov, and L.~M.~K. Vandersypen,
\newblock Science {\bf 318}, 1430 (2007).

\bibitem{NadjPerge2010}
S.~Nadj-Perge, S.~M. Frolov, E.~P. A.~M. Bakkers, and L.~P. Kouwenhoven,
\newblock Nature {\bf 468}, 1084 (2010).

\bibitem{Pioro2007}
M.~Pioro-Ladrière, Y.~Tokura, T.~Obata, T.~Kubo, and S.~Tarucha,
\newblock Appl. Phys. Lett. {\bf 90}, 024105 (2007).

\bibitem{PhysRevLett.96.047202}
Y.~Tokura, W.~G. van~der Wiel, T.~Obata, and S.~Tarucha,
\newblock Phys. Rev. Lett. {\bf 96}, 047202 (2006).

\bibitem{Yoneda2018}
J.~Yoneda et~al.,
\newblock Nature Nanotechnology {\bf 13}, 102 (2018).

\bibitem{PhysRevResearch.2.012006}
X.~Croot et~al.,
\newblock Phys. Rev. Research {\bf 2}, 012006 (2020).

\bibitem{PhysRevLett.99.246601}
E.~A. Laird et~al.,
\newblock Phys. Rev. Lett. {\bf 99}, 246601 (2007).

\bibitem{PhysRevLett.110.107601}
M.~Shafiei, K.~C. Nowack, C.~Reichl, W.~Wegscheider, and L.~M.~K. Vandersypen,
\newblock Phys. Rev. Lett. {\bf 110}, 107601 (2013).

\bibitem{Tosi2017}
G.~Tosi et~al.,
\newblock Nature Communications {\bf 8}, 450 (2017).

\bibitem{Petersson2012}
K.~D. Petersson et~al.,
\newblock Nature {\bf 490}, 380 (2012).

\bibitem{Kha2015}
A.~Kha, R.~Joynt, and D.~Culcer,
\newblock Appl. Phys. Lett. {\bf 107}, 172101 (2015).

\bibitem{Hu2012}
X.~Hu, Y.~X. Liu, and F.~Nori,
\newblock Phys. Rev. B {\bf 86}, 035314 (2012).

\bibitem{Beaudoin2016}
F.~Beaudoin, D.~Lachance-Quirion, W.~A. Coish, and M.~Pioro-Ladri{\`{e}}re,
\newblock Nanotechnology {\bf 27} 464003 (2016).

\bibitem{Elzerman2004}
J.~M. Elzerman et~al.,
\newblock Nature {\bf 430}, 431 (2004).

\bibitem{PhysRevLett.102.027601}
D.~R. McCamey, J.~van Tol, G.~W. Morley, and C.~Boehme,
\newblock Phys. Rev. Lett. {\bf 102}, 027601 (2009).

\bibitem{PhysRevB.100.125430}
M.~Benito et~al.,
\newblock Phys. Rev. B {\bf 100}, 125430 (2019).

\bibitem{Jang2020}
W.~Jang et~al.,
\newblock npj Quantum Information {\bf 6}, 64 (2020).

\bibitem{wang2017edsr}
Y.~Wang, C.-Y. Chen, G.~Klimeck, M.~Y. Simmons, and R.~Rahman,
\newblock arXiv:1703.05370  (2017).

\bibitem{Osika2021}
E.~N. Osika et~al.,
\newblock Unplublished  (2021).

\bibitem{Wang2016a}
Y.~Wang, C.-Y. Chen, G.~Klimeck, M.~Y. Simmons, and R.~Rahman,
\newblock Scientific Reports {\bf 6}, 31830 (2016).

\bibitem{Hileeaaq1459}
S.~J. Hile et~al.,
\newblock Science Advances {\bf 4} (2018).

\bibitem{PhysRevB.97.241401}
R.~Ferdous et~al.,
\newblock Phys. Rev. B {\bf 97}, 241401 (2018).

\bibitem{Kranz2020}
L.~Kranz et~al.,
\newblock Advanced Materials {\bf 32}, 1 (2020).

\bibitem{Weber2018}
B.~Weber et~al.,
\newblock npj Quantum Information {\bf 4}, 61 (2018).

\bibitem{PhysRevLett.103.110501}
F.~Motzoi, J.~M. Gambetta, P.~Rebentrost, and F.~K. Wilhelm,
\newblock Phys. Rev. Lett. {\bf 103}, 110501 (2009).

\bibitem{Struck2020}
T.~Struck et~al.,
\newblock npj Quantum Information {\bf 6}, 40 (2020).

\bibitem{Pla2013}
J.~J. Pla et~al.,
\newblock Nature {\bf 496}, 334 (2013).

\bibitem{PhysRevLett.100.067601}
J.~R. Petta et~al.,
\newblock Phys. Rev. Lett. {\bf 100}, 067601 (2008).

\bibitem{PhysRevLett.104.226807}
M.~Gullans et~al.,
\newblock Phys. Rev. Lett. {\bf 104}, 226807 (2010).

\bibitem{Theis_2018}
L.~S. Theis, F.~Motzoi, S.~Machnes, and F.~K. Wilhelm,
\newblock {EPL} (Europhysics Letters) {\bf 123}, 60001 (2018).

\bibitem{PhysRevA.82.042339}
E.~Lucero et~al.,
\newblock Phys. Rev. A {\bf 82}, 042339 (2010).

\bibitem{PhysRevX.2.011006}
L.~Trifunovic et~al.,
\newblock Phys. Rev. X {\bf 2}, 011006 (2012).

\bibitem{Viennot408}
J.~J. Viennot, M.~C. Dartiailh, A.~Cottet, and T.~Kontos,
\newblock Science {\bf 349}, 408 (2015).

\bibitem{Mi156}
X.~Mi, J.~V. Cady, D.~M. Zajac, P.~W. Deelman, and J.~R. Petta,
\newblock Science {\bf 355}, 156 (2017).

\bibitem{Samkharadze1123}
N.~Samkharadze et~al.,
\newblock Science {\bf 359}, 1123 (2018).

\bibitem{PhysRevApplied.5.044004}
N.~Samkharadze et~al.,
\newblock Phys. Rev. Applied {\bf 5}, 044004 (2016).

\bibitem{Veldhorst2014}
M.~Veldhorst et~al.,
\newblock Nature Nanotechnology {\bf 9}, 981 (2014).

\bibitem{Kane1998}
B.~E. Kane,
\newblock Nature {\bf 393}, 133 (1998).

\bibitem{Feher1959}
G.~Feher and E.~A. Gere,
\newblock Physical Review {\bf 114}, 1245 (1959).

\bibitem{Buch2013a}
H.~B{\"{u}}ch, S.~Mahapatra, R.~Rahman, A.~Morello, and M.~Y. Simmons,
\newblock Nature Communications {\bf 4}, 2017 (2013).

\bibitem{Watson2015}
T.~F. Watson, B.~Weber, M.~G. House, H.~B{\"{u}}ch, and M.~Y. Simmons,
\newblock Physical Review Letters {\bf 115}, 166806 (2015).

\bibitem{Petta2004}
J.~R. Petta, A.~C. Johnson, C.~M. Marcus, M.~P. Hanson, and A.~C. Gossard,
\newblock Physical Review Letters {\bf 93}, 1 (2004).

\bibitem{Kim2015}
D.~Kim et~al.,
\newblock Nature Nanotechnology {\bf 10}, 243 (2015).

\bibitem{Boross2016}
G.~S. {Peter Boross} and A.~P{\'{a}}lyi,
\newblock Nanotechnology {\bf 27} 314002 (2016).

\end{thebibliography}
\bibliographystyle{aip}

\appendix

\section{Conversion of the donor-donor Hamiltonian to the generic flopping-mode Hamiltonian}
\label{HamiltonianConversion}
We will show in the following that the full Hamiltonian describing the a double donor quantum dot system can be reduced to the four dimensional flopping-mode Hamiltonian in Eq.~\ref{eq:genHam} in the main text. This generalised Hamiltonian accurately describes the flopping-mode operation of the system but does not include leakage into nuclear spin states. The spin states in Eq.~\ref{eq:genHam} correspond to the combined electron-nuclear spin state of the phosphorus atoms, such that the electron flip-flops with the single nuclear spin $N_R = 1$ on the right dot while all other $N_L$ nuclear spins on the left dot do not participate in the dynamics.

The full Hamiltonian of the double quantum dot system with a total of $N = N_L + N_R$ nuclear spins can be written in the product basis $\left(\bigotimes_{k=1}^{N_L}\ket{\Uparrow^k/\Downarrow^k}\right)_L\otimes\ket{\Uparrow/\Downarrow}_R\otimes\ket{\uparrow/\downarrow}\otimes\ket{L/R}$ of the combined nuclear and electron spin as well as charge Hilbert spaces $\mathscr{H}_n$,  $\mathscr{H}_s$ and $\mathscr{H}_c$, respectively:

\begin{multline}
H=\gamma_e \pmb{B}\cdot \pmb{s}+\gamma_n\pmb{B}\cdot\sum_{k=1}^{N}\pmb{i}^k + \left(\epsilon\tau_z+t_c\tau_x\right)\\
+\sum_{k=1}^{N_L} A_{L,k}(\pmb{i}^k\cdot\pmb{s})\left(\mathds{1}+\tau_z\right)/2
+A_R\left(\pmb{i}^{N}\cdot\pmb{s}\right)\left(\mathds{1}-\tau_z\right)/2.
\label{eq:Ham1}
\end{multline}
Here we have defined the spin vector operators $\pmb{s}$ and $\pmb{i}^k$, of the electron and the k-th donor nucleus respectively, $\tau_i$ are the Pauli-operators acting on the charge subspace $\mathscr{H}_c$, $\pmb{B} = (0,0,B_0)$ is the external static magnetic field, $A_{L,k}$ is the $k$th contact hyperfine strength for the left quantum dot and $A_R$ is the hyperfine term for the right donor. Note that for convenience we have defined the right donor nuclear spin to be the $N$th nuclear spin operator.

The only coupling terms within the nuclear and electron spin subspace $\mathscr{H}_n\otimes\mathscr{H}_s$ are due to the hyperfine interaction. The full Hilbert space (electron, nuclear, and charge) can be decomposed into a direct sum of $H$-invariant subspaces according to their total spin polarisation $m$ (electron and nuclear spin),
\begin{equation}
\mathscr{H}=\bigoplus_{m=-(N+1)/2}^{(N+1)/2}\mathscr{H}_m^{N+1}=\bigoplus_{m=-(N+1)/2}^{(N+1)/2}\mathscr{H}_{s,m}^{N+1}\otimes\mathscr{H}_c.
\end{equation}
Note that the electron spin introduces the extra state (summation is over $N$ nuclear spins and 1 electron spin) and that the decomposition of the spin subspaces into $\mathscr{H}_{s,m}^{N+1}$ is carried over to the charge subspace. Due to spin conservation, the charge part of the Hamiltonian only connects states with the same subspace $\mathscr{H}_m^{N+1}$ of total spin $m$ and as a result simply doubles the size of the Hilbert space. Table~\ref{hilbertdimension} highlights the dimension of the invariant subspaces $\mathscr{H}_m^{N+1}$ of same spin polarisation $m$, for different donor numbers $N$.
\begin{table}[h!]
\centering
\caption{Dimensions of the invariant spin and charge subspaces of same spin polarisation $m$ with a single electron spin and $N$ donors.}
\begin{tabular}{|c|c|c|c|c|c|c|c|c|c|c|c|}
\hline
& \multicolumn{11} {c|} {$m$}\\
\hline
$N$ & -5/2 & -2 & -3/2 & -1 & -1/2 & 0 & 1/2 & 1 & 3/2 & 2 & 5/2\\
\hline
1 & 0 & 0 & 0 & 2 & 0 & 4 & 0 & 2 & 0 & 0 & 0\\
2 & 0 & 0 & 2 & 0 & 6 & 0 & 6 & 0 & 2 & 0 & 0\\
3 & 0 & 2 & 0 & 8 & 0 & 12 & 0 & 8 & 0 & 2 & 0\\
4 & 2 & 0 & 10 & 0 & 20 & 0 & 20 & 0 & 10 & 0 & 2\\
\hline
\end{tabular}
\label{hilbertdimension}
\end{table}
Any of the invariant subspaces in Table~\ref{hilbertdimension} offer the possibility of a flip-flop transition with the right nuclear spin except the two two-dimensional spaces $\mathscr{H}_{\pm (N+1)/2}^{N+1}$ that correspond to when the electron and nuclear spin(s) are fully polarised. The $N=1$ system (a single nuclear spin in the right quantum dot) corresponds to the quantum dot-donor (flip-flop) qubit and is the only case where one of subspace is four-dimensional and directly corresponds to a flopping-mode EDSR qubit. In all other values of $N$ the subspaces are larger then four-dimensional since the electron spin can flip-flop with more than one nuclear spin. In the donor-donor implementation in the main text ($N=3$, 2 nuclei on the left quantum dot and 1 on the right quantum dot) there are therefore 5 invariant subspaces with spin polarization $m= -2,\,-1,\,0,\,1,\, 2$ and respective dimensions $2,\,8,\,12,\,8,\,2$. The $m=\pm2$ subspaces correspond to all the spins being parallel: $\ket{\Downarrow \Downarrow \Downarrow \downarrow}$ and $\ket{\Uparrow \Uparrow \Uparrow \uparrow}$, respectively and cannot be used for EDSR since there is no electron-nuclear flip-flop transition. If the system reaches either of these states then NMR or dynamic nuclear polarisation would be needed to flip one of the nuclear spins into the opposite spin state. The $m=0$ subspace is especially attractive as the spectator nuclear spins on the left quantum dot can be initialised  within that subspace in such a way as to minimise the effective longitudinal magnetic field gradient as discussed extensively in the main text.

It is possible to reduce the Hamiltonian further, by treating the coupling to the $N_L$ nuclear spins perturbatively. Under the condition that the subspaces are non-degenerate, it is possible to fully remain within the qubit subspace by performing an appropriate state initialisation and by driving adiabatically at the frequency defined by the qubit splitting. The individual dipole moments and energy gaps all determine how fast a transition can be driven adiabatically, without leaking into the other states. The superconducting community has undertaken extensive work to design pulses sequences that reduce leakages to non-qubit subspaces while allowing fast driving, and thus minimise the influence of dephasing and relaxation errors. We will show in the qubit error section how we model the leakage out of the qubit subspace, and how we engineered the pulse shape to minimise the latter.

The Hamiltonian in Eq.~\ref{eq:Ham1} can be approximated by a first order Schrieffer-Wolff transform. Effectively, we restrict the Hamiltonian to the four dimensional subspace spanned by the spin states $\ket{N_L}\otimes\ket{\Downarrow}\otimes\ket{\uparrow}$ and $\ket{N_L}\otimes\ket{\Uparrow}\otimes\ket{\downarrow}$, and the two orbital charge states $\ket{L}$ and $\ket{R}$. The state $\ket{N_L}$ corresponds to the nuclear spin configuration of all $N_L$ nuclear spins in the left dot. This can be achieved by performing the following transformations on the Hamiltonian:
\begin{equation}
\pmb{i}^k\cdot\pmb{s}\mapsto\left\{
\begin{aligned}
&\frac{1}{4}\left(-\mathds{1}+2\sigma_x\right) &\text{if }k=N\\
&\langle \pmb{i}_z^k\rangle\sigma_z/2 & \text{if }k < N
\end{aligned}
\right. 
\end{equation}
where now $\sigma_i$ is defined in the new four-state basis. The nuclear Zeeman terms become:
\begin{equation}
\pmb{i}_z^k\mapsto\left\{
\begin{aligned}
&-\sigma_z/2 &\text{if }k=N\\
&\langle \pmb{i}_z^k\rangle\mathds{1} & \text{if }k < N
\end{aligned}
\right. 
\end{equation}
These transformations essentially select the matrix elements of the multidimensional matrices $\pmb{i}^k\cdot\pmb{s}$ and $\pmb{i}_z^k$ that correspond to the last two dimensions of the Hilbert space (right nuclear spin state and electron spin state).

After performing the transformation and subtracting global energy shifts, we get:
\begin{multline}
\label{thequantHf}
H_E=\left[\frac{1}{2}\left(\left(\gamma_e+\gamma_n\right)B_z+M_n\right)\sigma_z+\frac{A_R}{4}\sigma_x\right]\\
+\left[\left(\epsilon+\frac{A_R}{8}\right)\tau_z+t_c\tau_x\right] +\frac{1}{4}\left(2 M_n\sigma_z-A_R\sigma_x\right)\tau_z,
\end{multline}
where we capture the influence of the effective magnetic field produced from the spectator nuclear spins as the averaged hyperfine interaction $M_n=\sum_{k=1}^{N_L} A_{L,k} \langle \pmb{i}_z^k\rangle/2$.

We can diagonalize the spin-like terms ($\sigma_i$) in Eq.~\ref{thequantHf}, which results in a small rotation of the quantisation axis due to the nuclear spin Zeeman and hyperfine terms. Afterwards, we finally recover the Hamiltonian of the form described in Eq.~\ref{eq:genHam}, with the following parameters:

\begin{align}
\Omega_z&=\sqrt{\Omega_s^2+\left(\frac{A_R}{2}\right)^2},\\ 
\text{with }\Omega_s &=\left(\gamma_e+\gamma_n\right) B_0 + M_n+\frac{\delta \Omega^{(2)}}{2},\\
\epsilon_A&=\epsilon+\frac{A_R}{8}+\frac{\delta \Omega^{(2)}}{4},\\
\Delta \Omega_{\parallel}&=  \left(2 M_n-\delta \Omega^{(2)}\right)\text{cos}(\theta)-A_R\text{sin}(\theta),\\
\Delta \Omega_{\perp}&=  A_R\text{cos}(\theta) +\left(2 M_n-\delta \Omega^{(2)}\right)\text{sin}(\theta),
\end{align}
The correction term, $\delta \Omega^{(2)}=O\left(\frac{A_L^2}{(\gamma_e + \gamma_n) B_0}\right)$ arises from the higher order terms of the Schrieffer-Wolff approximation which we neglect for the following analysis since they only have a small effect on the Hamiltonian parameters.Very close to nuclear spin level crossings, some even higher order effects describing nuclear spin state hybridisation via the electron hyperfine interaction become relevant, but can safely be neglected by staying clear of the levels crossings during driving of the qubit, and can be traversed diabatically when initialising the qubit.

The angle $\theta$ corresponds to a very small rotation of the qubit quantisation axis due the perpendicular component of the hyperfine interaction:
\begin{align}
\text{cos}(\theta)&=\frac{\Omega_s}{\Omega_z}\approx 1,\\
\text{sin}(\theta)&=\frac{A_R/2}{\Omega_z}\approx 0.
\end{align}
Finally, the new spin basis is defined as,
\[\tilde{\uparrow}/\tilde{\downarrow}=\frac{1}{\sqrt{2}}\left( \mp\sqrt{1\pm\text{cos}(\theta)},\mp \sqrt{1\mp\text{cos}(\theta)} \right), \]
expressed in the explicit combined nuclear and electron spin basis $\{\ket{N_L}\otimes\ket{\Downarrow\uparrow},\ket{N_L}\otimes\ket{\Uparrow\downarrow}\}.$

Note that similarly to the quantum dot-donor qubit, the coupling between the qubit states is purely determined by the hyperfine coupling to the nuclear spin that the electron spin flip-flops with ($A_R$). However, $\Delta \Omega_{\parallel}$ is determined by the averaged hyperfine interaction $M_n$ of the electron with the nuclear spins in the left quantum dot, which are not involved in the qubit dynamics, and that we therefore call the spectator nuclear spins. As we covered in the main text, we can engineer this averaged hyperfine interaction $M_n$ in order to minimise $\Delta \Omega_{\parallel}$ and in turn increase the dephasing time of the qubit.

\section{Adiabatic orbital state transfer}
The adiabatic orbital state transfer displayed in Fig.~\ref{fig:2}~d) is calculated numerically for a 2P-1P device operated at the (2,1) $\leftrightarrow$ (3,0) electron state with the nuclear spins on the left quantum dot initialised in antiparallel spin states at a magnetic field of $B=0.3\,\text{T}$ and a tunnel coupling of $t_c=5.9\,\text{GHz}$. We chose a difference in the hyperfine coupling to the two nuclei in the left dot of $\Delta A_L = 1\,\text{MHz}$ based on the measured couplings from a 2P quantum dot~\cite{Hileeaaq1459}. We start the adiabatic ramp at $\epsilon(t=0)=110\,\text{GHz}$ away from the charge degeneracy point where the spin-like state only has a 0.1\% of charge component and qubit coherence times are approximately those of a single electron spin. At this position, we initialise the qubit into an even superposition of the two qubit states, $\ket{0} \equiv \ket{\Downarrow \Uparrow \Uparrow \downarrow -}$ and $\ket{1}  \equiv \ket{\Downarrow \Uparrow \Downarrow \uparrow -}$. We then perform a numerical time evolution of that state under the influence of a linear detuning pulse ending at $\epsilon=0$ where the qubit can be driven electrically. At the end of the pulse of duration $t_p$, some of the qubit population has leaked out of the qubit subspace. The leakage probability into the charge excited states is calculated by summing the end state population in the excited charge states $\ket{+}$, whereas the leakage probability due to nuclear spin states on the left quantum dot flipping is estimated by summing the end state population in the nuclear spin states in the ground charge states $\ket{-}$ (excluding the qubit states).

\section{Theoretical error model for the flopping-mode EDSR qubit}
\begin{figure}
\begin{center}
\includegraphics[width=1\columnwidth]{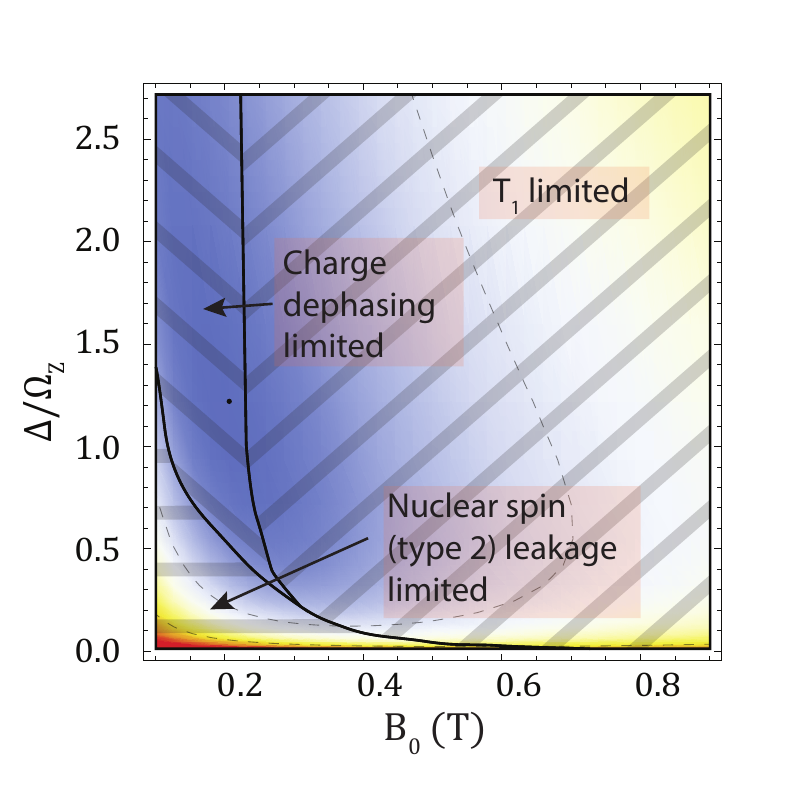}
\end{center}
\vspace{-0.5cm}
\caption{{\bf Limiting error source for the 2P-1P qubit at the (2,1)$\leftrightarrow$(3,0) transition with $\Delta A_L= 1\,\text{MHz}$.} 
Overlayed over the error plot from figure~\ref{fig:3}~b), we show the three regions where different errors sources dominate the total error at the optimal drive amplitude.  For high spin-charge detuning $\Delta/E_z$ and high magnetic field, the $T_1$ error limits the total error. For low magnetic fields, charge dephasing mostly dominates the error. Leakage errors only start being significant for small spin charge detuning and magnetic fields. In that region, only leakage to the near degenerate states is significant.}
\vspace{-0.5cm}
\label{fig:DominatingErrorSource}
\end{figure}

During electric driving of the qubit, dephasing, $T_1$ relaxation and state leakage introduce errors in the operation of the qubit. 
In our error model, we include dephasing of the qubit due to electric field noise, $T_1$ relaxation of the charge qubit, and leakage out of the qubit states. We do not include pure spin dephasing ($\sim$kHz) and relaxation ($\sim$Hz) as both are orders of magnitude lower than the charge related error sources~\cite{Muhonen2014}. In Fig.~\ref{fig:DominatingErrorSource}, we display the dominating error sources corresponding to the error calculation in Fig.~\ref{fig:3} of the main text. At low magnetic field and hence low tunnel coupling the charge $T_1$ relaxation is small and the qubit error is dominated by dephasing and leakage errors. At low spin-charge detunings, $\Delta = 2 t_c - \Omega_z$ the qubit is limited by leakage due to unwanted nuclear spin flips on the left quantum dot. At high magnetic field and large spin-charge detuning excited charge state $T_1$ relaxation dominates the qubit error. In the following sections we describe the different error sources associated with the donor-donor flopping-mode qubit that were investigated to generate Fig.~\ref{fig:DominatingErrorSource}.

\subsection{Flopping-mode EDSR Hamiltonian with electric drive} 
Driving of the qubit is achieved by applying an electric field burst $E(t)=E_{d}\cdot g(t,t_p) \cos(\omega_{d}t)$, oscillating with frequency $\omega_d$, for a pulse time $t_p$ and with a time dependent pulse envelope $g(t,t_p)$. In all our simulations, we use a Gaussian pulse shape depicted in Fig.~\ref{fig:2}~e) (top graph) which has been shown to reduce excited state leakage when driving superconducting transmon qubits~\cite{PhysRevLett.103.110501},
\begin{align}
g(t,t_p)=\frac{1-\text{exp}\left(\frac{2 t(t_p-t)}{t_p^2}\right)}{1-e^{1/2}}.
\label{GaussianPulseShape}
\end{align}
The symmetric Gaussian pulse shapes cannot fully reduce leakage during qubit driving but can help reverse leakage state excitation (see Fig.~\ref{fig:2}~e) (bottom graph)). 

The oscillating electric field drive can be written as a time dependent detuning parameter in the flopping mode Hamiltonian in equation~\ref{eq:genHam}:
\[\tilde{\epsilon}(t)=\epsilon+\epsilon_d(t),\] where $\epsilon_d(t):=\frac{e E_d d}{2 h} g(t, t_p) \cos(\omega_d t)$ is the detuning drive, $\epsilon$ is the static detuning and $d$ is the distance between the two quantum dots.

The full driven system can then be expressed as the sum of the static Hamiltonian in Eq.~\ref{eq:genHam} and the time-dependent drive Hamiltonian $H_d=\epsilon_d(t) \tau_z $ expressed in the basis defined by the product states $\{\ket{i},\,i=0,\dots,\,3\}=\left\{\ket{\downarrow -},\ket{\uparrow -}, \ket{\downarrow +},\ket{\uparrow +}\right\}$ the driven Hamiltonian takes the form:
\begin{equation}
\label{flopping4levelmatrix}
H_{rl}=\begin{pmatrix}
0 &\Omega_r &\Omega_l & 0\\[0.3em]
\Omega_r &\Omega_s &0&\Omega_l\\[0.3em]
 \Omega_l &0 &\Omega_c&\Omega_r \\[0.3em]
 0& \Omega_l & \Omega_r  &\Omega_c+\Omega_s
 \end{pmatrix},
 \end{equation}
where $\Omega_s/\Omega_c$ is the spin/charge qubit energy ($\Omega_c = 2\sqrt{\epsilon^2 + t_c^2}$) respectively, where $\Omega_r$ is the coupling between the two qubit states, and $\Omega_l$ is the coupling of each qubit state to it's corresponding excited charge states. In the following section we use this Hamiltonian in  Eq.~\ref{flopping4levelmatrix} to estimate the charge dephasing error for the donor-donor flopping-mode EDSR qubit. As was described in Section~\ref{HamiltonianConversion}, this Hamiltonian describes the system very well apart from nuclear spin leakage.

%

\subsection{Charge dephasing error modelling}

To model the charge dephasing error of the qubit we assume that the charge noise couples through small perturbations $\delta\epsilon$ in the detuning $\epsilon$. We further assume that these perturbations are well described as a random variable $\delta\epsilon$ described by a Gaussian probability distribution function $P(\delta\epsilon)$ , centred about the value of $\epsilon$~\cite{Kranz2020} with a standard deviation of $\sigma\epsilon$ . For comparison with the other flopping-mode EDSR proposals we use an electric field noise of about $125\,\text{V/m}$, similar to that used in other flopping mode proposals~\cite{Tosi2017, PhysRevB.100.125430} and corresponding to a standard deviation in the static detuning parameter of about $\sigma\epsilon = 0.3\,\text{GHz}$.

The charge dephasing error of the unitary evolution associated with the $\pi/2 - X$ gate is determined by the deviation of the expectation value of the noisy unitary evolution projected onto the ideal unitary evolution $U_{\text{id}}$ of an initial qubit state $\Psi_{i,\delta\epsilon}$~\cite{Tosi2017}, averaged over the charge noise detuning distribution, $P(\delta\epsilon)$:

\begin{align}
\label{dephasingformula}
\text{e}_\epsilon&=1-\left\langle\left|\bra{\Psi_{i,\delta\epsilon}}U_{\delta\epsilon}^\dagger U_{\text{id}}\ket{\Psi_{i,\delta\epsilon}}\right|^2\right\rangle_{\delta\epsilon}.
\end{align}


We have developed an analytical model of the state overlap $O(\delta\epsilon,\Psi_{i,\delta\epsilon}):=\left|\bra{\Psi_{i,\delta\epsilon}}U_{\delta\epsilon}^\dagger U_{\text{id}}\ket{\Psi_{i,\delta\epsilon}}\right|^2$ allowing for averaging of the error over all possible initial states $\Psi_{i,\delta\epsilon}$ of the Bloch sphere, which is crucial considering that the gate error can vary by up to an order of magnitude depending on the initial qubit state. 

\begin{figure}
\begin{center}
\includegraphics[width=1\columnwidth]{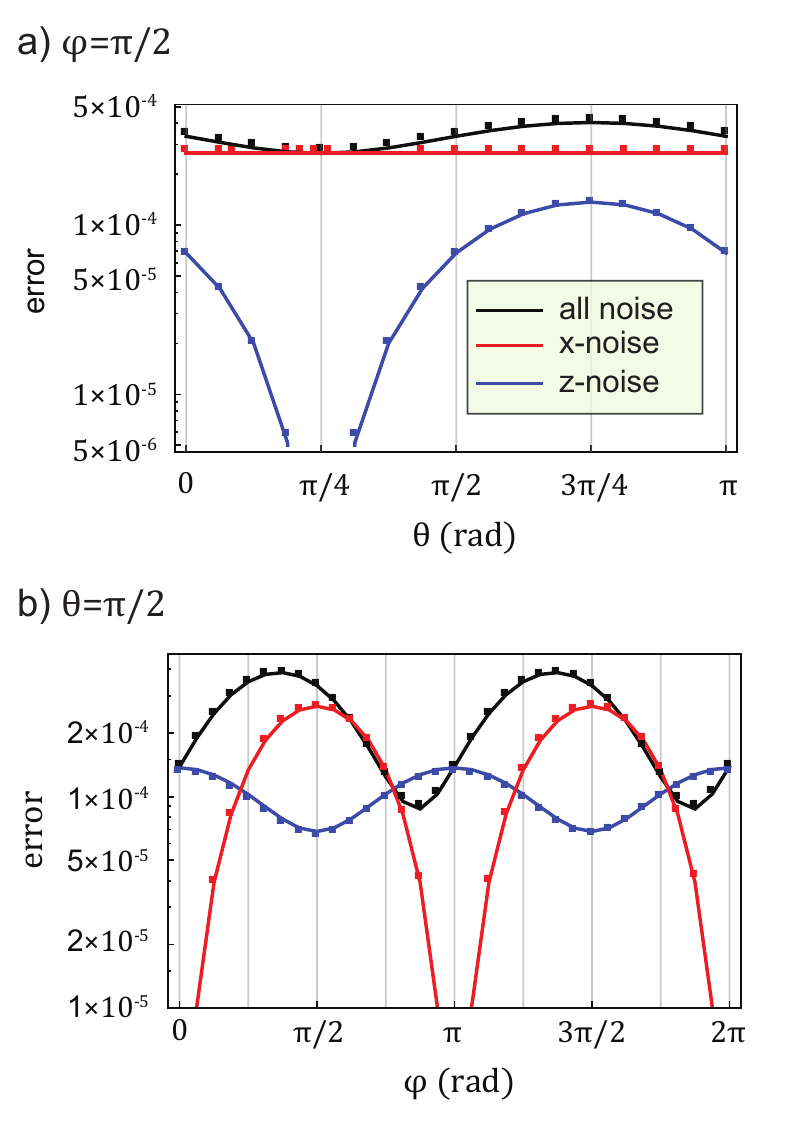}
\end{center}
\vspace{-0.5cm}
\caption{{\bf Charge noise dephasing modelling.} Both {\bf a)} and {\bf b)} show the angle dependence of the dephasing charge noise for $\phi=\pi/2$ and $\theta=\pi/2$  as a function of the longitudinal and azimuthal angles $\theta$ and $\phi$ respectively. The dark line displays the error when considering all channels through which charge noise can couple into the system. The red/blue lines only consider the x-/z- charge noise channels.
 The analytical model (lines) accurately fits the numerical calculation (square markers). 
 In a), the z-error goes to zero at $\theta=\pi/4$ because variations $\delta z$ are echoed out when passing the pole.
 In b), the x-error goes to zero at $\phi=0\,(\text{mod} \, \pi)$, because the start state is on the x-axis of the Bloch sphere.
 We used a magnetic field of $0.3\,\text{T}$ at $\epsilon=0$ and $t_c=4.5\,\text{GHz}$, for a drive amplitude $\epsilon_d=0.2\,\text{GHz}$.
 }
\vspace{-0.5cm}
\label{fig:dephasingmodel}
\end{figure}

Using the above model, charge noise effectively couples into the noisy unitary time evolution $U_{\delta\epsilon}$ through the unwanted perturbation of the different Hamiltonian parameters in Eq.~\ref{flopping4levelmatrix}. Provided the system is driven adiabatically, the dynamics are mostly confined to the qubit subspace which is well described by the two-level Hamiltonian, $\Omega_z \sigma_z +\Omega_r\sigma_x$, where $2\Omega_z$ is the qubit energy splitting and $\Omega_r$ the qubit Rabi frequency. Both $\Omega_z$ and $\Omega_r$ are dependent on $\epsilon$ and offer distinct pathways for charge noise to couple into the time evolution, which we define as the $z-/x-$ noise channels, respectively. For a given detuning perturbation $\delta \epsilon$ we write the instantaneous values as $\Omega_z(\epsilon +\delta \epsilon)=\Omega_z(\epsilon)+\delta z$ and $\Omega_r(\epsilon +\delta \epsilon)/2=\Omega_r(\epsilon )/2 +\delta x=x+\delta x$. In the rotating frame, when driving the qubit on resonance the reduced two-level Hamiltonian becomes:
\begin{equation}
\label{redHam}
H_r(\delta z, \delta x)=\delta z\, \sigma_z + (x + \delta x)\,\sigma_x.
\end{equation}
The time evolution associated with this Hamiltonian, can be modelled analytically if we approximate the Gaussian drive pulse $\epsilon_d\,g(t,t_p)$ as a constant pulse $\epsilon_d \,\bar{g}$, where $\bar{g}=0.633$ is the average value of $g(t,t_p)$. With the Hamiltonian, Eq.~\ref{redHam} and drive pulse, we calculate the state overlap $O(\delta\epsilon,\Psi_{i,\delta \epsilon})$ in Eq.~\ref{dephasingformula} for an initial state $\Psi_{i}=\cos{(\theta/2)}\ket{0}+\sin{(\theta/2)}e^{\imath \phi}\ket{1}$ using,
\begin{multline}
\label{simplifiedOverlap}
O(\delta\epsilon,\Psi_{i,\delta \epsilon})\approx \left|\bra{\Psi_i} U(H_r(0,0),t_{\pi/2})^\dagger\right. \\
\left. \cdot U(H_r(\delta z, \delta x),t_{\pi/2}) \ket{\Psi_i}\right|^2.
\end{multline}
Here we estimate the $\pi/2$ gate time to be $t_{\pi/2}=\pi/(4\bar{g} x)$. The unitary time evolution operator $U$ of the Hamiltonian $H$ can be calculated explicitly as the matrix exponential $U(H,t_g)=\exp\left(-\imath H t_g\right)$. In Fig.~\ref{fig:dephasingmodel} we compare the fully numerical error calculation (square markers) with the analytical error model described above (solid line) for a range of initial states on the Bloch sphere. The numerical calculation computes the overlap in Eq.~\ref{dephasingformula} using the full flopping mode Hamiltonian~\ref{flopping4levelmatrix}, while the analytical model uses the analytical expression corresponding to Eq.~\ref{simplifiedOverlap}.
In black we include all the noise channels present in Eq.~\ref{flopping4levelmatrix}, whereas in red and blue we only include the $x-/z-$ noise channels. The analytical model can be seen to fit the numerical calculations very well, highlighting the fact that the charge noise predominantly enters the dephasing error through the $z$- and $x$- dephasing channels that affect the qubit states directly, and that dephasing contributions through the excited charge leakage states can be neglected.

Figure~\ref{fig:dephasingmodel} also highlights certain qubit states that are protected against noise channels within the qubit subspace.  This translates into a large variation in error rate depending on the initial qubit state, motivating the need for averaging the qubit error. In Fig.~\ref{fig:dephasingmodel}~a) the $z-$ error goes to zero for the initial state with $\phi=\pi/2$ and $\theta=\pi/4$. This initial state has no $z-$dephasing as it corresponds to a symmetric rotation through the $\ket{1}$ state of the Bloch sphere such that any unwanted phase accumulation during the first half of the pulse is reversed in the second half of the pulse. Two other states where the $x-$dephasing approaches zero are shown in Fig.~\ref{fig:dephasingmodel}~b) at $\theta=\pi/2$ and $\phi=0\,(\text{mod} \, \pi)$. These angles correspond to the two initial states along the $x-$axis of the Bloch sphere. These two states are not affected by $x-$rotations and consequently do not experience dephasing due to noise along the $x-$axis. The inclusion of both errors (in black) limits the magnitude of the error variations in this instance; however, there is still a significant variation in error rate depending on the initial state that needs to be considered when operating the qubit. Additionally, if a particular qubit is dominated by either $z-$ or $x-$ noise then the variation in error as a function of the initial state can vary by orders of magnitude.

The first step towards calculating the full dephasing error is to analytically integrating the overlap model in Eq.~\ref{simplifiedOverlap} over all the initial states on the Bloch sphere. We find that the state averaged overlap between the perturbed (noisy) and the non-perturbed (ideal) time evolution is then given by:
\begin{multline}
\label{blochaverageddeph}
\left\langle O(\delta\epsilon,\Psi_{i,\delta\epsilon})\right \rangle_\mathscr{B} \approx O_\mathscr{B}(\delta z, \delta x)=\left.\frac{1}{6\Omega_{2L}^2}\right( 4(x+\delta x)^2+3\delta z^2\\
\left.+\delta z^2 \cos{(\frac{\pi}{2} \frac{\Omega_{2L}}{x})}+2(x +\delta x)\Omega_{2L}\sin{(\frac{\pi}{2} \frac{\Omega_{2L}}{x})}\right),
\end{multline}
where we have defined the Rabi splitting $\Omega_{2L}=\sqrt{\delta z^2 +(x+\delta x)^2}$. As expected the expression evaluates to 1 for $\delta z=0$ and $\delta x=0$, since the noisy time evolution is equal to the ideal (noiseless) case, that is, there is no charge noise in the system. 

In our case both x and z noise perturbations $\delta z$ and $\delta x$ are dependent on the electric detuning noise variable $\delta\epsilon$. 
The second and final step in obtaining the fully averaged analytical charge dephasing error is performed by averaging $1- O_\mathscr{B}(\delta z(\delta\epsilon),\delta x(\delta\epsilon))$ (as described in eq. \ref{blochaverageddeph})  over the the electric detuning noise variable $\delta\epsilon$ :
\begin{equation}
\left\langle \text{e}_\epsilon\right \rangle_\mathscr{B}=1-\left\langle O_\mathscr{B}(\delta z(\delta\epsilon),\delta x(\delta\epsilon))\right\rangle_{\delta\epsilon}.
\label{FullAnalyticalErrorFormula}
\end{equation}
We calculate this average over the Gaussian distributed random variable $\delta\epsilon$ numerically.

In the next section, we investigate the various leakage pathways present in the donor-donor implementation. The leakage errors become dominant for strong qubit driving and for near degeneracies in the hyperfine couplings of the electron to the different phosphorus nuclear spins.

\subsection{Leakage modelling}
\label{sect:leakagemodelling}
\begin{figure}
\begin{center}
\includegraphics[width=1\columnwidth]{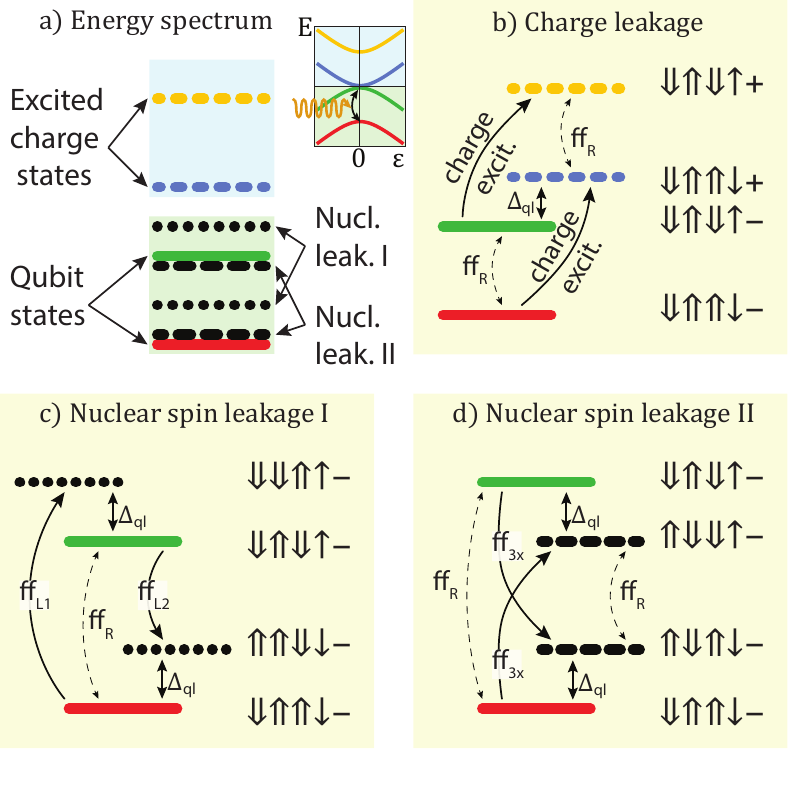}
\end{center}
\vspace{-0.5cm}
\caption{{\bf  Leakage pathways for the 2P-1P donor-based flopping-mode EDSR qubit.} {\bf a)} Simplified energy spectrum at $\epsilon = 0$ (see inset) for the flopping-mode qubit in the main text. The qubit states, $\ket{0} \equiv \ket{\Downarrow\Uparrow\Uparrow\downarrow-}$ and $\ket{1} \equiv \ket{\Downarrow\Uparrow\Downarrow\uparrow-}$ are shown in red and green, respectively. The excited charge states of the same electron and nuclear spin states as the qubit states are shown in blue and yellow. The black dotted (leakage type I) and dashed (leakage type II) lines correspond to the different nuclear spin leakage states discussed in the main text. {\bf b)} The charge excitation leakage pathway. Charge leakage, shown by the solid arrow lines, results from accidental excitation of the charge state of the double quantum dot. The qubit frequency is shown as ff$_{R}$ (that is a flip-flop transition with the right nuclear spin and the leakage state energy separation, $\Delta_{ql}$ used in simulating the leakage error is shown between the green and blue states. {\bf c)} Type I nuclear spin leakage corresponds to a single flip-flop transition of the electron with the first nuclear spins on the left quantum dot, ff$_{L,1}$ ($\ket{\Downarrow\Uparrow\Downarrow\uparrow-} \rightarrow \ket{\Uparrow\Uparrow\Downarrow\downarrow-}$) and the second nuclear spin on the left quantum dot, ff$_{L,2}$ ($\ket{\Downarrow\Uparrow\Uparrow\downarrow-} \rightarrow \ket{\Downarrow\Downarrow\Uparrow\uparrow-}$) shown by the black arrows. {\bf d)} Type II nuclear spin leakage occurs when the electron flip-flops with all three nuclear spins simultaneously, ff$_{3\times}$. Flip-flop transitions of this type can occur in both directions ($\ket{\Downarrow\Uparrow\Uparrow\downarrow-} \rightarrow \ket{\Uparrow\Downarrow\Downarrow\uparrow-}$ and $\ket{\Downarrow\Uparrow\Downarrow\uparrow-} \rightarrow \ket{\Uparrow\Downarrow\Uparrow\downarrow-}$) and cause leakage out of the computational basis of the qubit.}
\vspace{-0.5cm}
\label{fig:leakage}
\end{figure}

The second error type that we consider in our model is state leakage of the qubit subspace. 
The donor-donor qubit states defined in the main text can potentially leak to the 10 other states of the Hilbert space of same magnetisation, see Fig.~\ref{fig:leakage}a).
Leakage into any of the 6 states in the excited charge state branches (light blue square in Fig.~\ref{fig:leakage}~a) and in the inset) is dominated by the direct charge excitation ($\ket{-} \rightarrow \ket{+}$) from the qubit states shown in red and green in Fig.~\ref{fig:leakage}, to their excited charge state counterparts in blue and yellow, which have the same electron and nuclear spin configuration as the qubit states. Leakage to these two excited charge states during electric driving is dominant leakage process due to their large electric-dipole moment. Leakage into the excited charge subspace will be referred to as the ``charge leakage pathway" and is represented in Fig.~\ref{fig:leakage}~b). 
In the ground charge state branch (light green square in Fig.~\ref{fig:leakage}~a) and in the inset), there are four states that the qubit can leak into, depicted by black dotted lines in Fig.~\ref{fig:leakage}. These four states can be broken into two more leakage pathways that we will reference to as ``nuclear spin leakage pathways". The first nuclear spin leakage pathway corresponds to a flip-flop transition of the electron with one of the nuclear spin of the left quantum dot instead of the right dot (see Fig.~\ref{fig:leakage}~c) ). Indeed, the ground (excited) qubit state $\ket{\Downarrow\Uparrow\Uparrow\downarrow-}$ ($\ket{\Downarrow\Uparrow\Downarrow\uparrow-}$) can leak to the spin state $\ket{\Downarrow\Downarrow\Uparrow\uparrow-}$ ($\ket{\Uparrow\Uparrow\Downarrow\downarrow-}$) via a flip-flop transition, $\text{ff}_{L2}$ ($\text{ff}_{L1}$) with the second (first) nuclear spin on the left quantum dot. We call this leakage pathway ``type I nuclear spin leakage". The second nuclear spin leakage pathway in the donor-donor qubit corresponds to leakage from the qubit states into the near degenerate levels $\ket{\Uparrow\Downarrow\Downarrow\uparrow-}$ and $\ket{\Uparrow\Downarrow\Uparrow\downarrow-}$ via 3 simultaneous electron-nuclear flip-flop transitions with all the nuclear spins in the system ($\text{ff}_{3\times}$). This second nuclear spin pathway is displayed in Fig.~\ref{fig:leakage}~d), and will be referred to as ``type II nuclear spin leakage".

For all three independent leakage pathways (one charge and two nuclear spin flips) the four level system consisting of the qubit states $\ket{0}$ and $\ket{1}$ and their respective leakage state ($\ket{2}$ and $\ket{3}$) is described by the Hamiltonian in the basis $\{\ket{i}, \,i=0,\dots,\, 3\}$,
\begin{equation}
\label{leakage4levelmatrix}
H_{rl}=\begin{pmatrix}
0 &\Omega_r/2 &\Omega_l/2 & 0\\[0.3em]
\Omega_r/2 &0 &0&\Omega_l/2\\[0.3em]
 \Omega_l/2 &0 &\Delta_{ql}&\Omega_{rl}/2 \\[0.3em]
 0& \Omega_l/2 & \Omega_{rl}/2 &\pm\Delta_{ql}
 \end{pmatrix}.
 \end{equation}
We define $\Omega_r$ to be the coupling between the two qubit states, $\Omega_l$ the coupling to the leakage states, $\Omega_{rl}$ the coupling between the leakage states, and $\Delta_{ql}$ the energy gap between qubit and leakage states. The coupling strength $\Omega_{rl}$ between the leakage states and the sign of the gap $\pm\Delta_{ql}$ turn out to be irrelevant to the total leaked state proportion due to the coupling strengths $\Omega_l$ being symmetrical. Using the Hamiltonian in Eq.~\ref{leakage4levelmatrix} we can model the different leakage pathways analytically and substitute in the various strengths of the coupling and detuning terms.

To minimise state leakage we adiabatically drive the qubit transition by slowly increasing and then decreasing the drive amplitude in time using a symmetric Gaussian pulse shape displayed at the top of Fig.~\ref{fig:2}~e). The time-dependent drive leads to a time-dependent occupation of the leakage states in all three leakage pathways that increases and decreases with the pulse amplitude. The use of symmetric continuous pulse shape allows for most of the leakage state population (both charge and nuclear spin leakage) to be de-excited in the second half of the pulse~\cite{PhysRevLett.103.110501}, for small drive amplitudes less than the energy separation between the qubit state and the leakage state (see Fig.~\ref{fig:2}~e) ). We call the integrated leakage population during the pulse ``reversible leakage" as it is mostly reversed at the end of the pulse.

Pulse shaping however cannot fully reverse the leakage population. We call the remaining leakage population at the end of the pulse ``irreversible leakage". It is a source of error for all leakage pathways as it leads to a finite probability of the system to be measured outside of the qubit subspace. The irreversible leakage error is simply given by the occupation probability of the leakage states at the end of the pulse.
The reversible leakage mechanism can also lead to errors if the leakage state is itself prone to errors. We have seen in the previous section that charge dephasing via the excited charge states is negligible. The same holds true for the nuclear leakage states. However, relaxation of the leakage state can lead to significant errors for the charge leakage pathway. Indeed, the excited charge states can relax to the ground state due to $T_1$ charge relaxation.  The excited charge state is temporarily occupied during qubit operation leading to a finite probability for the qubit state to relax back to the ground state. We call this drive-$T_1$ error as it only occurs during driving of the qubit. Reversible leakage into nuclear spin states (in the ground charge state branch) does not lead to additional relaxation errors because all nuclear spin states have long relaxation times. 

Reversible leakage can be characterised by the integrated probability of the qubit state being in the two leakage states, during the $\pi/2$ Gaussian pulse of duration $t_{\pi/2}$, with the aim to later use the quantity in order to calculate the $T_1$ relaxation error associated with it:
\begin{equation}
\label{iddef}
I_{d}:=\int_0^{t_{\pi/2}} \sum_{i=2}^{3}\left|\braket{\Psi(t')}{i}\right|^2 \mathrm{d}t'.
\end{equation}
In the following section~\ref{T1section}, we will derive how this leakage integral $I_{d}$ in eq.~\ref{iddef} enters the calculation of the drive-$T_1$ error.
The integral, $I_{d}$ can be estimated by assuming a noiseless unitary time evolution of an initial state on the Bloch sphere. We find that the integral is independent of the start state and can be well approximated to second order in $\frac{\Omega_l}{\Delta_{ql}}$):
\begin{align}
\label{idformula}
I_{d}&\approx\alpha_d \frac{1}{\Omega_r}\frac{\Omega_l^2}{\Delta_{ql}^2},
\end{align}
The coefficient $\alpha_d$ is related to the Gaussian pulse shape used to drive the qubit and is equal to 0.046 for the specific case described in Eq.~\ref{GaussianPulseShape}. The integral is independent of the initial qubit state due to the fact that the coupling strengths, $\Omega_l$ of the qubit states to the leakage states are equal so that any superposition of the two qubit states is equally likely to leak out of the qubit subspace. The total leakage state population is inversely proportional to the coupling $\Omega_r$ between the qubit states and is thus proportional to the gate time $t_{\pi/2}=\pi/(2\bar{g} \Omega_r)$ reflecting the fact that shorter pulses lead to a smaller integrated leakage probability. The leakage is also inversely proportional to the qubit-leakage state energy gap, highlighting that smaller energy separations lead to larger leakage probabilities.
As we will cover in the following section~\ref{T1section}, this analytical model in eq~\ref{idformula}, is used in the calculation of the $T_1$ relaxation error.

We now turn to the irreversible leakage error which is the probability of the system being in the leakage states $\ket{2}$ and $\ket{3}$ at the end of the $\pi/2$~pulse,
\begin{equation}
e_\text{leak}=p_\text{leak}=\sum_{i=2}^{3} \left| \braket{\Psi(t_{\pi/2})}{i}\right|^2.
\end{equation}
The leakage probability, $p_\text{leak}$ has two distinct regimes depending on the respective magnitude of the qubit driving strength $\Omega_r$ and the energy gap $\Delta_{ql}$ to the nearest leakage state. In both regimes the leakage probability is related to the ratio $\lambda=\Omega_l/\Omega_r$ of the leakage and qubit coupling strength.
In the first regime, where the qubit drive amplitude $\Omega_r$ is smaller then the energy gap to the nearest leakage state $\Delta_{ql}$ (``weak driving regime"), the leakage population grows polynomially with drive amplitude $\Omega_r$ as the qubit drive become less adiabatic:
\begin{equation}
\label{equ:leakagepopulationformula}
p_\text{leak}\approx\alpha_\text{leak}\lambda^2 \frac{\Omega_r^4}{\Delta_{ql}^4},
\end{equation}
where $\alpha_\text{leak}=0.37$ is a constant related to the Gaussian pulse shape determined through numerical simulation. For the charge leakage pathway, a significant leakage contribution is attributed to the factor $\lambda^2$ in \ref{equ:leakagepopulationformula}, because the coupling of the excited charge state is always greater than the qubit state, and typically results in a factor $\lambda$ much larger then 1. However, for this charge leakage pathway, the gap $\Delta_{ql}$ is usually much larger then the qubit coupling, so that the remaining factor $\left(\frac{\Omega_r}{\Delta_{ql}}\right)^4$ is much smaller then unity, and the leakage probability can remain small despite a large ratio $\lambda$.
In the second regime, in which the qubit drive amplitudes becomes larger or comparable to the energy gap to the leakage state $\Omega_r >~\Delta_{ql}$ (``Strong driving regime") the leakage population asymptotically approaches a constant value. Indeed, at high drive amplitudes, the power-broadened qubit transition overlaps with the leakage transition and both transition are driven. If the coupling to the leakage state is smaller then the coupling between the qubit states ($\lambda<1$) the qubit will only leak out at a maximum probability described only by the ratio $\lambda$:
\begin{equation}
\label{equ:asymptoticleakagepopulation}
p_\text{leak} \approx \left(\frac{\pi}{4}\right)^2 \lambda^2.
\end{equation}
The leakage probability  for the nuclear spin leakage of type II can easily fall into this regime, because the energy gap to the leakage state ($\Delta_{ql}=\Delta A_L/2\simeq 500\,\text{kHz}$), is often larger then the optimal Rabi frequency. However, despite this small energy gap, the leakage probabilty in this particular leakage pathway remains small. Indeed, the coupling strength $\Omega_l$ to the leakage state in this leakage pathway is much smaller then the qubit coupling, leading to a factor $\lambda$ much smaller then one and resulting into a low leakage probability according to eq.~\ref{equ:asymptoticleakagepopulation}.

For the error calculations in the main text we use a combination of Eq.~\ref{equ:leakagepopulationformula} and Eq.~\ref{equ:asymptoticleakagepopulation} to model the leakage probability within each leakage pathway when driving the qubit. In the next section we will cover the how $T_1$ charge relaxation can lead to two types of errors, one related to the excited charge state proportion naturally present in the qubit, the other linked to the excited charge state proportion excited during the reversible leakage process that was described in this section.

\subsection{Charge $T_1$ relaxation modeling}
\label{T1section}
Relaxation errors of the proposed donor-donor qubit can be due to nuclear spin, electron spin or charge relaxation. Any relaxation of the electron spin, the nuclear spin or the excited charge state translates into relaxation of the qubit. These three relaxations occur over a wide range of characteristic timescales.

The $T_1$ relaxation times of the nuclear spin of a phosphorus donor in silicon has been measured to be of the order of minutes~\cite{Kane1998,Pla2013} whereas the relaxation time of electron spins on phosphorus donor quantum dots has been measured to be of the order of seconds~\cite{Feher1959,Buch2013a,Watson2015} at magnetic fields of about 1 T. The relaxation time of a charge qubit defined by the symmetric and antisymmetric superposition of two tunnel coupled quantum dot orbitals however has been measured to be of the order of only a few nanoseconds in GaAs quantum dots~\cite{Petta2004} and in Si/SiGe gate-defined quantum dots~\cite{Kim2015}. The charge relaxation rate $1/T_1^c$ in silicon donor quantum dots has been theoretically estimated for a charge qubit defined by a phosphorus donor quantum dot and an interface quantum dot. Boross~et.~al.~\cite{Boross2016} predict the charge relaxation rate to be proportional to the charge qubit energy splitting and to the square of the tunnel coupling $2t_c$ between the two quantum dots,
\begin{equation}
\label{chargeT1}
1/T_1^c\approx\Theta \left(2\sqrt{\epsilon^2 +t_c^2}\right)\cdot(2 t_c)^2\,\, \text{(GHz)},
\end{equation}
where the coefficient $\Theta\approx 2.37\times 10^{-6}~(\text{ns}^2)$ is a silicon specific constant~\cite{Boross2016,Tosi2017}.
At zero detuning, where the qubit is operated, the charge relaxation rate is proportional to $t_c^3$. For a typical charge qubit splitting of $11\,\text{GHz}$ corresponding to a magnetic field of $0.4\,\text{T}$ for an electron spin qubit, equation~\ref{chargeT1} yields a relaxation time of about $300\,\text{ns}$. Since the electron and nuclear spin relaxation times in our qubit can be expected to be of the order of seconds or even minutes we expect that charge relaxation will be the dominating relaxation mechanism. In our calculations, we will use Eq.~\ref{chargeT1} to calculate the relaxation rate $1/T_1^c$ of the pure charge qubit.

The charge T1 relaxation if the charge excited state is well described by an exponential decay process described by the error:  $\frac{1}{2}(1-\exp{(-t/T_1^c)})$. This error does not fully describe the relaxation of the qubit state since it only partially overlaps with the excited charge state, making it less probable for the qubit to decay in the equivalent time as the charge qubit. The exponential decay of our proposed qubit therefore needs to include the time-integrated overlap of the qubit wave function with the excited charge state. The qubit relaxation error can be calculated using~\cite{Tosi2017},
\begin{align}
\label{T1fullerrorformula}
\text{e}_{T_1}=\frac{1}{2}\left(1-\text{Exp} \left[-\int_0^{t_{\pi/2}}\sum_s \left| \braket{\Psi(t')}{s,+} \right|^2 \frac{1}{T_1} \mathrm{d}t' \right]\right),
\end{align}
where $\ket{s,+}=\ket{s}\otimes\ket{+}$ are the product states containing the excited charge states, $\ket{+}$. The qubit relaxation errors grow exponentially with the gate time $t_{\pi/2}$ and with the overlap $\sum_s \left| \braket{\Psi(t)}{s,+} \right|^2$ of the qubit states with the excited charge state.

There are two ways by which the qubit states can overlap with the excited charge state during a $\pi/2-X$ gate. Firstly, while the qubit ground state $\ket{0}$ does not overlap at all with the excited charge state (due to the large energy separation, $\propto t_c$), the qubit excited state $\ket{1}$ is engineered to have a small excited charge state proportion $p_{1,+}=\sum_s\left|\braket{1}{s,+}\right|^2$. This is a result of the hybridisation of the spin qubit with the charge qubit that allows electric driving of our qubit. 
Secondly, the qubit states can also overlap with the excited charge state by reversible leakage into the excited charge states during qubit operation. 
Those two effects result in a total time dependent overlap of the qubit states with the excited charge state given by:
\begin{multline}
\label{T1decomp}
\sum_s \left| \braket{\Psi(t)}{s,+} \right|^2 \approx \left|\braket{\Psi(t)}{1}\right|^2 p_{1,+} + \sum_{i=2}^3\left|\braket{\Psi(t)}{i}\right|^2
\end{multline}
The relaxation error in Eq.~\ref{T1fullerrorformula} is related to the time integral of this overlap~\ref{T1decomp}:
\begin{align}
\int_0^{t_{\pi/2}}\sum_s \left| \braket{\Psi(t')}{s,+} \right|^2 \mathrm{d}t'&=I_1 \cdot p_{1,+}+ I_d,
\end{align}
where $I_{1}:=\int_0^{t_{\pi/2}} \left|\braket{\Psi(t')}{1}\right|^2 \mathrm{d}t'$, and the integral $I_{d}\approx\alpha_d \frac{1}{\Omega_r}\frac{\Omega_l^2}{\Delta_{ql}^2}$ was derived in the previous section describing reversible leakage (Eq.~\ref{idformula}) and describes the excited charge leakage population. It is dependent on the Rabi frequency $\Omega_r$, the coupling strength to the excited state $\Omega_l$ and the energy separation $\Delta_{ql}$ between the qubit states and the nearest excited charge state.

The integral $I_1$ of the $\ket{1}$ state overlap can be approximated by calculating the noiseless time evolution of an initial qubit state $\ket{\Psi_i}=\cos{\theta/2}\ket{0}+\sin{\theta/2}e^{\imath \phi}\ket{1}$ during a $\pi/2-X$ gate, 
\begin{align}
I_{1}(\theta,\phi)&\approx \frac{1}{\Omega_r}\frac{1}{4}\left(\pi-2 \cos{\theta}-2\sin{\theta}\sin{\phi}\right).
\end{align}
The full relaxation error in Eq.~\ref{T1fullerrorformula} for a given initial state can then be written as,
\begin{align}
\text{e}_{T_1}(\theta,\phi)=\frac{1}{2}\left(1-e^{-I_1 (\theta,\phi)p_{1,+}/T_1}  e^{-I_d/T_1}\right).
\end{align}

Finally, the relaxation error averaged over the Bloch sphere is given by:
\begin{align}
\label{finalrelaxationerror}
\left\langle \text{e}_{T_1}\right\rangle_\mathscr{B}=\frac{1}{2}\left(1-\left\langle e^{-I_1 (\theta,\phi)p_{1,+}/T_1} \right\rangle_\mathscr{B} e^{-I_d/T_1}\right).
\end{align}
The Bloch sphere average of the term $e^{-I_1(\theta,\phi) p_{1,+}/T_1}$ can be approximated analytically. Integration over $\phi$ results in a Bessel function which can be approximated to third oder in $\beta=\frac{p_{1,+}}{\Omega_r T_1}$,
\begin{equation}
\label{averagedidleterm}
\left\langle e^{-I_i (\theta,\phi)O_{1,+}/T_1} \right\rangle_\mathscr{B} \approx e^{\frac{1}{4}(2 +\pi)\beta}\frac{\beta-2 +e^\beta(\beta +2)}{4\beta}.
\end{equation}
The relaxation error of the qubit is calculated using Eq.~\ref{finalrelaxationerror} and Eq.~\ref{averagedidleterm}, and the parameters entering the equation are calculated numerically. The estimation of the pure charge relaxation rate uses Eq.~\ref{chargeT1}. 

\subsection{Combining all errors}
Finally, we combine the dephasing error, the relaxation error and the irreversible leakage errors into one total error formula, assuming that these errors originate form independent random processes:
\begin{equation}
\text{e}_{\text{tot}}(\theta,\phi)=1-(1-\text{e}_{\epsilon} (\theta,\phi))(1-\text{e}_{T_1}(\theta,\phi)(1-\text{e}_{\text{leak}}).
\end{equation}

The average of this error over the Bloch sphere can be approximated as the product of the averages of each error, yielding the final error metric used in the main text: 
\begin{align}
\left\langle \text{e}_{\text{tot}}(\theta,\phi)\right\rangle_\mathscr{B} &\approx \left\langle 1-\text{e}_{\epsilon}\right\rangle_\mathscr{B}  \left\langle 1-\text{e}_{T_1} \right\rangle_\mathscr{B} (1-\text{e}_{\text{leak}}).
\end{align}

\section{Calculation of the spin-cavity coupling and the qubit dephasing time}
We investigate the qubit-cavity coupling characteristic, which is shown in the Fig.~\ref{fig:4} of the main text. Strong coupling of a cavity to a qubit can be achieved if the qubit-cavity coupling strength, $g_{sc}$ is larger then the dephasing rate $\gamma$ of the qubit as well as the decay rate $\kappa$ of the cavity. The coupling strength, $g_{sc}$ can be calculated as the product of the qubit electric dipole transition matrix element $\chi_{01}$ and the electric field amplitude produced by the cavity at the location of the qubit. Following the cavity simulation of Osika~\emph{et al.}~\cite{Osika2021}, we use detuning amplitudes of about $\epsilon_c=100\,\text{MHz}$, and a cavity decay rate $\kappa=1\,\text{MHz}$. The detuning amplitude corresponds to zero point voltage fluctuations of the cavity of the order of $0.4\,\mu\text{V}$ for quantum dots separated by about $10\,\text{nm}$, or equivalently to cavity electric fields of about $10\text{V/m}$. We calculate the transition matrix element $\chi_{01}$ numerically and estimate the qubit dephasing rate, $\gamma = 1/T_2^*$ by converting the average qubit error using the formula,
\begin{equation}
T_2^*\approx2\sqrt{2}\sqrt\frac{t_{\pi/2}^2}{\text{Log}\left(\frac{1}{1-2 \,\text{error}}\right)}.
\end{equation}
The dephasing rate is then calculated as a function of magnetic field strength and tunnel coupling, while the  cavity detuning amplitude $\epsilon_c$ and the cavity decay rate $\kappa$ are assumed to be constant across the parameter range investigated in Fig.~\ref{fig:4}.

\end{document}